\documentclass[onecolumn]{aastex62}

\usepackage{amsmath}

\accepted{May 9, 2023}
\submitjournal{ApJ}

\shorttitle{AASTeX v6.3.1 Sample article}
\shortauthors{\c{S}ahin et al.}
\graphicspath{{./}{figures/}}
\def\ion#1#2{#1$\;${\small\rm\@{#2}}\relax}
\newcommand{\uat}[2]{\href{http://astrothesaurus.org/uat/#2}{#1 (#2)}}

\begin{document}


\title{Spatial and Temporal Analysis of Quiescent Coronal Rain over an Active Region}


\correspondingauthor{Seray \c{S}ahin}
\email{seray.sahin@northumbria.ac.uk}

\author[0000-0003-3469-236X]{Seray \c{S}ahin}
\author[0000-0003-1529-4681]{Patrick Antolin}
\affiliation{Department of Mathematics, Physics and Electrical Engineering, Northumbria University, Newcastle Upon Tyne, NE1 8ST, UK}
\author[0000-0001-5315-2890]{Clara Froment}
\affiliation{LPC2E, CNRS/University of Orl\'eans/CNES, 3A avenue de la Recherche Scientifique, Orl\'eans, France}
\author[0000-0002-7451-9804]{Thomas A. Schad}
\affiliation{National Solar Observatory, 22 \'{}\={O}hi\'{}a K\={u} Street, Pukalani, HI 96768, USA}

\begin{abstract}

The solar corona produces coronal rain, hundreds of times colder and denser material than the surroundings. Coronal rain is known to be deeply linked to coronal heating, but its origin, dynamics, and morphology are still not well understood. The leading theory for its origin is thermal instability (TI) occurring in coronal loops in a state of thermal non-equilibrium (TNE), the TNE-TI scenario. Under steady heating conditions, TNE-TI repeats in cycles, leading to long-period EUV intensity pulsations and periodic coronal rain.  In this study, we investigate coronal rain on the large spatial scales of an active region (AR) and over the long temporal scales of EUV intensity pulsations to elucidate its distribution at such scales.  We conduct a statistical study of coronal rain observed over an AR off-limb with IRIS and SDO imaging data, spanning chromospheric to transition region (TR) temperatures. The rain is widespread across the AR, irrespective of the loop inclination, and with minimal variation over the 5.45-hour duration of the observation. Most rain has a downward ($87.5\%$) trajectory; however, upward motions ($12.5\%$) are also ubiquitous. The rain dynamics are similar over the observed temperature range, suggesting that the TR and chromospheric emission are co-located on average. The average clump widths and lengths are similar in the SJI channels and wider in the AIA 304 channel. We find ubiquitous long-period EUV intensity pulsations in the AR. Short-term periodicity is found (16 min) linked to the rain appearance, which constitutes a challenge to explain under the TNE-TI scenario.  
\end{abstract}


\section{Introduction} \label{sec:intro}

\begin{figure*}[ht]
\includegraphics[width=\textwidth]{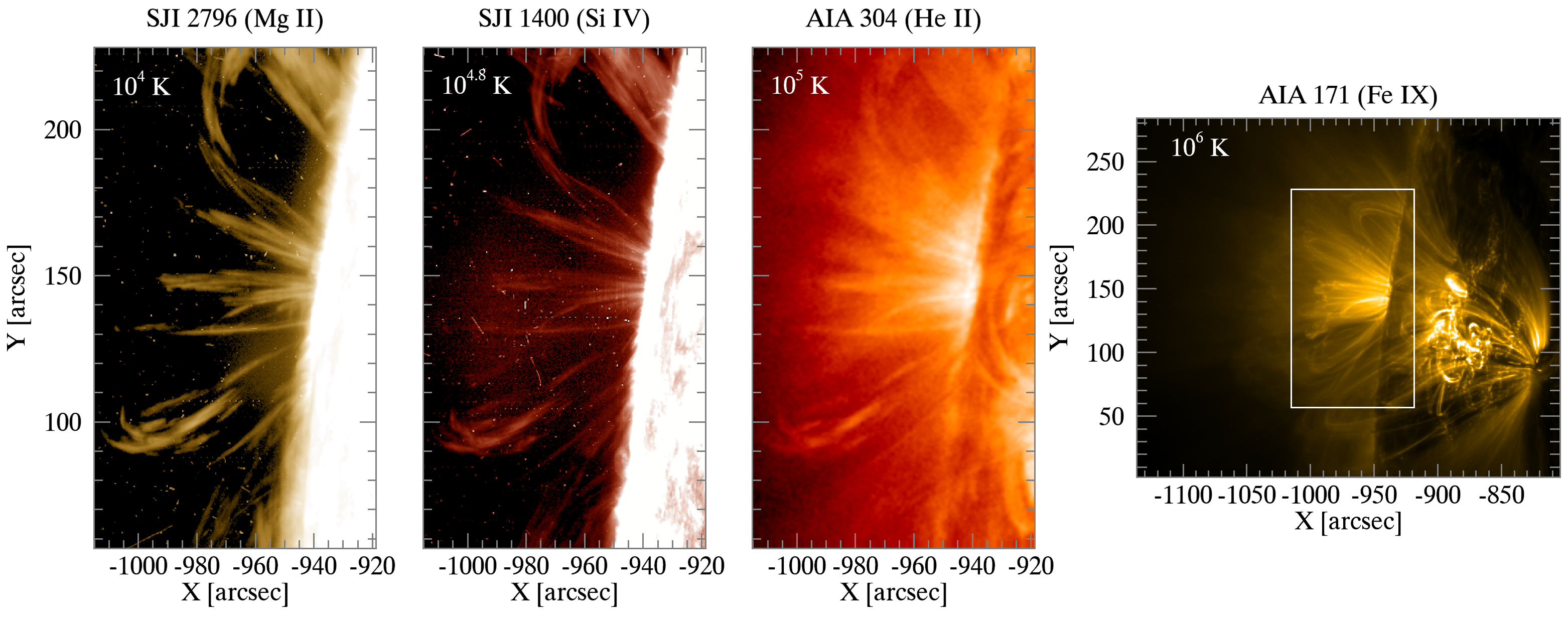}
\caption{The studied active region at the East limb of the Sun on 2 June 2017,  observed by IRIS/SJI 1400~\AA (left), SJI 2796~\AA (second from left) and SDO/AIA 304~\AA~(third from left). The separated panel on the right shows a wider FOV with SDO/AIA 171~\AA over the same region,  with the white rectangle outlining the FOV shown on the left panels. All images were obtained summing over 10 images in the interval 07:28:00 UT - 07:34:28 UT. (Movie of this figure is available. The movie shows off-limb coronal rain seen in the SJI 2796~\AA, SJI 1330~\AA, and AIA 304~\AA~falling towards the solar surface along coronal loop structures all around a sunspot that sits in the middle of the FOV. The rain is seen to fall continuously over the observation time of 5.45 hours, starting from 07:28 UT to 12:55 UT.)}
\label{figure1}
\end{figure*}

The solar corona, in addition to being hot, tenuous and diffuse, also hosts cold, dense and clumpy plasma structures such as prominences and coronal rain. This rain corresponds to partially ionized plasma that is commonly observed in active regions. Its temperature and density vary between 10$^{3} – 10^{5}$ K and $\approx 10^{10} – 10^{12}~$cm$^{-3}$, respectively. Coronal rain is best observed off-limb with high contrast against the dark background, where it appears in emission in chromospheric lines such as H$\alpha$, \ion{Ca}{II} H $\&$ K, and \ion{He}{I} \citep{DeGroof2004,Schad2018ApJ...865...31S,froment2020}, and in transition region lines such as \ion{Si}{IV}~1402~\AA\, or/and in \ion{He}{II}~304~\AA~\citep{Vashalomidze_2015AA...577A.136V,antolin2015}. It forms in a couple of minutes and falls toward the solar surface over an order of magnitude longer timescales along loop-like trajectories \citep{antolin2012}. The distribution of rain clump velocities are broad, from $10-200$~km~s$^{-1}$ with a mean value around 70~km~s$^{-1}$ \citep{schrijver2001,muller2005,antolin2012}, and are characterised by being lower than free-fall, which is generally accepted to be caused by gas pressure forces \citep{Oliver2014}. Another candidate is the ponderomotive force from transverse MHD waves, but \citet{Verwichte2017A&A...598A..57V} have shown that the wave amplitudes that are observed usually cannot account for the rain dynamics.

Coronal rain is believed to originate from thermal instability (TI) within a coronal loop in a state of thermal non-equilibrium (TNE), known as the TNE-TI scenario \citep{antolin2020, Antolin_Froment_10.3389/fspas.2022.820116} in which radiative cooling locally overcomes the coronal heating input \citep{muller2003,muller2004}. Although the exact role of TI is still under debate \citep{Klimchuk_2019SoPh..294..173K}, the TNE onset conditions are usually met in coronal loops with strongly stratified and high-frequency heating  \citep{Antiochos_1999ApJ...512..985A}. Numerical simulations allow to study the parameter space of TNE-TI and coronal rain formation. The results can then be compared with observations, which can then constrain the parameter space of the heating \citep{fang2013,fang2015,Froment2018ApJ...855...52F, Klimchuk_2019ApJ...884...68K, johnston2019}. The specific spatio-temporal heating distribution requirements to generate coronal rain, therefore, constitute stringent proxies of coronal heating. When the spatial and temporal distribution of the heating in a coronal loop are stable, TNE manifests through periodic cycles of global heating and cooling due to the inability of the loop to reach thermal equilibrium \citep{Kuin_1982AA...108L...1K}. The cooling through the coronal temperatures manifests as long-period EUV pulsations \citep{Auchere2014A&A...563A...8A,froment2015}. 





Coronal rain is observed mainly in three different forms in the solar atmosphere \citep{Antolin_Froment_10.3389/fspas.2022.820116}. The most common kind is termed quiescent coronal rain and is the main target of this study. It is observed in active region loops and is unrelated to visible flares. Another kind is flare-driven coronal rain and corresponds to the cool material observed in flare loops at the last stage of their cooling \citep{scullion2016}. The last kind is known as hybrid prominence/coronal rain complexes, which are recently studied structures forming in magnetic dips over null point topologies \citep{Liu_2016SPD....47.0402L,LiL_2018ApJ...864L...4L,Mason2019ApJ...874L..33M,ChenH_2022AA...659A.107C}.


On the morphological side, lengths and widths of coronal rain blobs vary depending on the temperature of formation of the observed wavelength, with widths of a few hundred km in a chromospheric line such as H$\alpha$ and close to  1000~km in a transition region line such as AIA~304~\AA, although a strong influence of spatial resolution is also expected. Lengths vary more strongly, with values up to  tens of Mm, and a dependence on temperature  is unclear \citep{antolin2012,antolin2015}. This clumpy and filamentary morphology of the rain is one of the most puzzling features, and is thought to be caused not only due to the interaction with the environment \citep[for example, through shear flows,][]{fang2015}, but also and particularly due to the nature of TI \citep{Waters_2019}. Recently, through 2.5D radiative MHD simulations, \citet{Antolin2022A} have shown that TI plays an important role in the observed variability and filamentary structure of the corona in the UV and EUV because of the impact of the rain onto the chromosphere, the strong compression downstream of the rain showers and since a thin but strongly emitting Condensation Corona Transition Region (CCTR) is created by TI, not only confined to the condensation region but extending along the loop. It has therefore been argued that coronal rain provides us with direct physical insight into the thermodynamics within coronal loops \citep{antolin2020}.


A major characteristic of coronal rain is that the clumps appear at very similar times over a relatively wide volume, and fall along a similar trajectory. This characteristic has led to the concept of a rain shower \citep{antolin2012}. Using the same dataset as the one for this work, \citet{Sahin2022ApJ...931L..27S} has conducted the first statistical analysis of showers. They observed widespread showers over the same active region as studied in the present paper, successfully identifying coronal loops.  Using the traced rain showers, they quantified the total coronal volume in the AR in the TNE state to be around $4.56\times10^{28}$ cm$^3$, that is, on the same order as the AR volume, with strong implications for the spatial and temporal coronal heating conditions at AR level. 

Contrary to ground-based observations, IRIS observations allow conducting relatively long-duration investigations of coronal rain over the same region. 
Using this IRIS capability, in this study, we present the first spatial and temporal statistical analysis of quiescent coronal rain over an entire active region with a significant time duration of 5.45~hours. Such duration is on the same order as long period intensity pulsations (2 to 16 hours \citep{Auchere2014A&A...563A...8A}) and therefore allows us to test and improve previous statistics of coronal rain properties, as well as its relation to the EUV corona through long-period intensity pulsations. The rest of the paper is organized as follows. In Section \ref{sec:data}, we introduce our dataset. The methodology and results are described in Section \ref{sec:method} and Section \ref{sec:result}, respectively. Finally, in Section \ref{sec:sum}, we summarize and discuss our observational results of the coronal rain.

\begin{figure*}[ht]
\includegraphics[width=\textwidth]{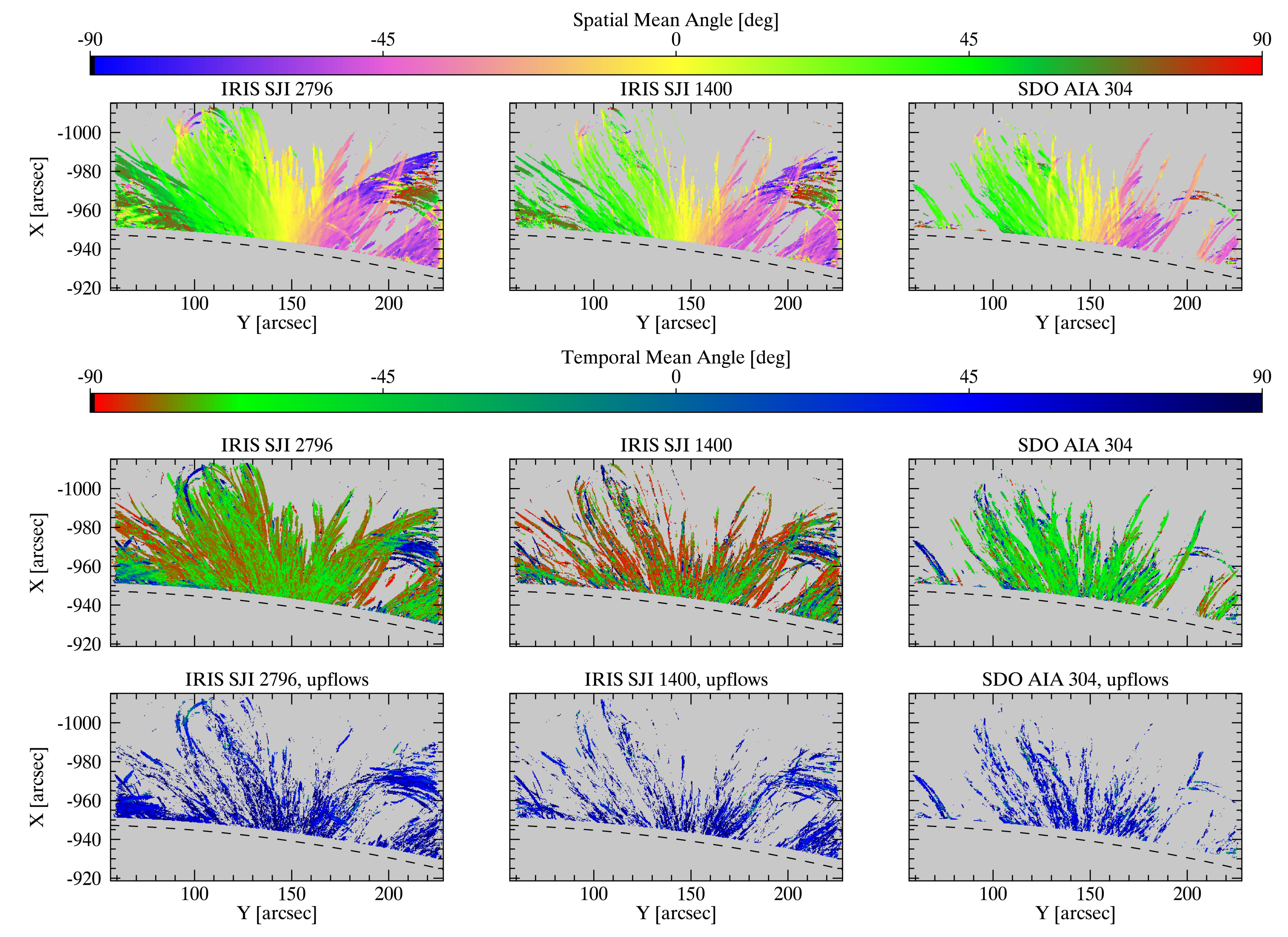}
\caption{Top: Average spatial mean angle maps over the entire time sequence as derived using the Rolling Hough Transform method. Middle and bottom: Average temporal mean angle maps separated into upward and downward (middle) and upward (bottom) motions. The spatial mean angle maps indicate the inclination of the rain with respect to the vertical direction, while temporal mean angle maps indicate the dynamical change along a trajectory. The dashed black curves indicate the solar limb.\label{mean_angles}}
\end{figure*}

\section{Observations} \label{sec:data}

The active region (AR) under study was observed by IRIS on 2017 June 2 between 07:28:00 UT – 12:54:39 UT when it was close to the East limb of the Sun (Figure~\ref{figure1}). We use level 2 slit-jaw imager (SJI) data from the Interface Region Imaging Spectrograph \citep[IRIS;][]{depontieu2014} combining two IRIS passbands; far-UV (FUV) and near-UV (NUV) passbands. These two passbands are centered on 1400~\AA\ (dominated by the \ion{Si}{IV}~1402.77~\AA~transition region line forming at $10^{4.8}$~K)  and on 2796~\AA~(dominated by the \ion{Mg}{II} 2796.35~\AA\ chromospheric line forming at 10$^{4}$~K), respectively. IRIS\footnote{\url{https://www.lmsal.com/hek/hcr?cmd=view-event\&event-id=ivo://sot.lmsal.com/VOEvent\%23VOEvent_IRIS_20170602_072800_3620110460_2017-06-02T07:28:002017-06-02T07:28:00.xml}} conducted a 64-step raster centered at [-970\arcsec,143\arcsec], providing a large maximum SJI field-of-view (FOV) of 232\arcsec$\times$182\arcsec. The pixel scale for this dataset is 0\arcsec.3327 pixel$^{-1}$, the average cadence for these SJI passbands are 43.1~s (SJI 1400) and 32.3~s (SJI 2796), and the exposure time for these channels is 15~s. 

We further analyse observations of this AR by the Atmospheric Imaging Assembly \citep[AIA;][]{lemen2012} on board of the Solar Dynamics Observatory  \citep[SDO;][]{pesnell2012} with 12~s cadence and an exposure time of 2.9~s. 
The AIA observations contain seven broad passbands, however, we focused on the AIA~304~\AA\ channel dominated by \ion{He}{II} line (forming at $\approx10^{5}$ K) for the rain detection and its analysis. The pixel size of AIA is 0\arcsec.6 pixel$^{-1}$, however, for the purpose of co-alignment, we have rebinned the AIA data to match the SJI plate scale. 

In order to distinguish between the different datasets used in this study, we will use specific names for each. The dataset we mentioned above hereafter will be referred to as the ``rain dataset", while the dataset below will be referred to as the ``pulsation dataset." We also study the long-term intensity pulsations of this region using 3-day (starting from the 1st of June and ending on the 4th of June) level-1.5 datasets in all seven AIA EUV channels (94~\AA, 131~\AA, 171~\AA, 193~\AA, 211~\AA, 304~\AA, and 335~\AA), focusing on a FOV of 332\arcsec$\times$282\arcsec~(see Figure \ref{figure1} right panel) FOV centered at [-970\arcsec,142\arcsec] (see Fig.~\ref{figure1}). For this study, we use a cadence of 5 minutes. The AIA data are read and calibrated to level 1.5 using the routine \textit{aia\_prep} from the Interactive Data Language SolarSoftware library (IDL SSW). The intensities are first normalised to the exposure times, and a 4$\times$4 pixel binning is applied to increase the signal-to-noise ratio. Please see \citet{froment2015,froment2020} for further details. Contrary to previous studies, we also check for the existence of short-term periodicities ($<$1 hour) in the same region.



\section{Methodology} \label{sec:method}

\subsection{Data preparation}
For the co-alignment between the AIA and SJI channels, we first resample the AIA data to the SJI plate scale. The co-aligning is then achieved by matching the solar limb and on-disk features seen in co-temporal images of AIA~1600~\AA~and SJI~2832~\AA, which form under similar conditions. Once the $x$ and $y$ shifts are obtained at various time instances, an interpolation is performed for any other time sequence and applied to all other channels. 

As discussed in \citet{Wulser2018SoPh..293..149W}, IRIS observations are influenced by a modest level of scattered light that is pointing-dependent and often difficult to characterize, especially for off-limb observations. Due to this scattered light and imperfect flat fielding, the rastering mode of IRIS results in a background contribution that shows apparent motion relative to the solar image and thereby causes a non-trivial, time-dependent noise in the SJI images that periodically increases the intensity in the image. This artificial periodic intensity is 17.2 min in both SJI 1400~\AA~and 2796~\AA, and it is more pronounced in SJI~2796 closer to the limb but is clearer relative to the coronal rain intensity high above the limb in SJI~1400~\AA. We partially corrected for this noise pattern by calculating the minimum intensity evolution over a raster for each pixel in both SJI channels. The SJI cube is then divided by the minimum intensity value at each pixel and time step.

The AIA~304~\AA~channel has a temperature response peak at $\approx10^5 K$ from \ion{He}{II}~304~\AA~emission. However, the bandpass also incorporates a secondary peak at $\approx10^{6.2} K$ due to \ion{Si}{XI}~303.32~\AA. When observing coronal rain off-limb, the intensity of both components can be similar because the Si XI component coming from the surrounding diffuse hot corona is greater in extent along the line-of-sight than the He II component coming from the small, clumpy coronal rain \citep{antolin2012,froment2020}. Besides temperature, both plasma emissions also differ strongly in their morphology. While the cool emission is clumpy, the hot emission is diffuse. We, therefore, use a morphology-based approach called Blind Source Separation (BSS) to separate both components, as outlined in \citet{Dudok-de-Wit:2013uw}. We have tested the combination of several AIA channels as source components with different numbers of morphological components, and we find that the combination of four source AIA channels (335~\AA, 304~\AA, 211~\AA, and 193~\AA) with 3 main morphological components is enough to isolate the clumpy coronal rain structure from a given 304 image.


\begin{figure*}[ht]
\includegraphics[width=\textwidth]{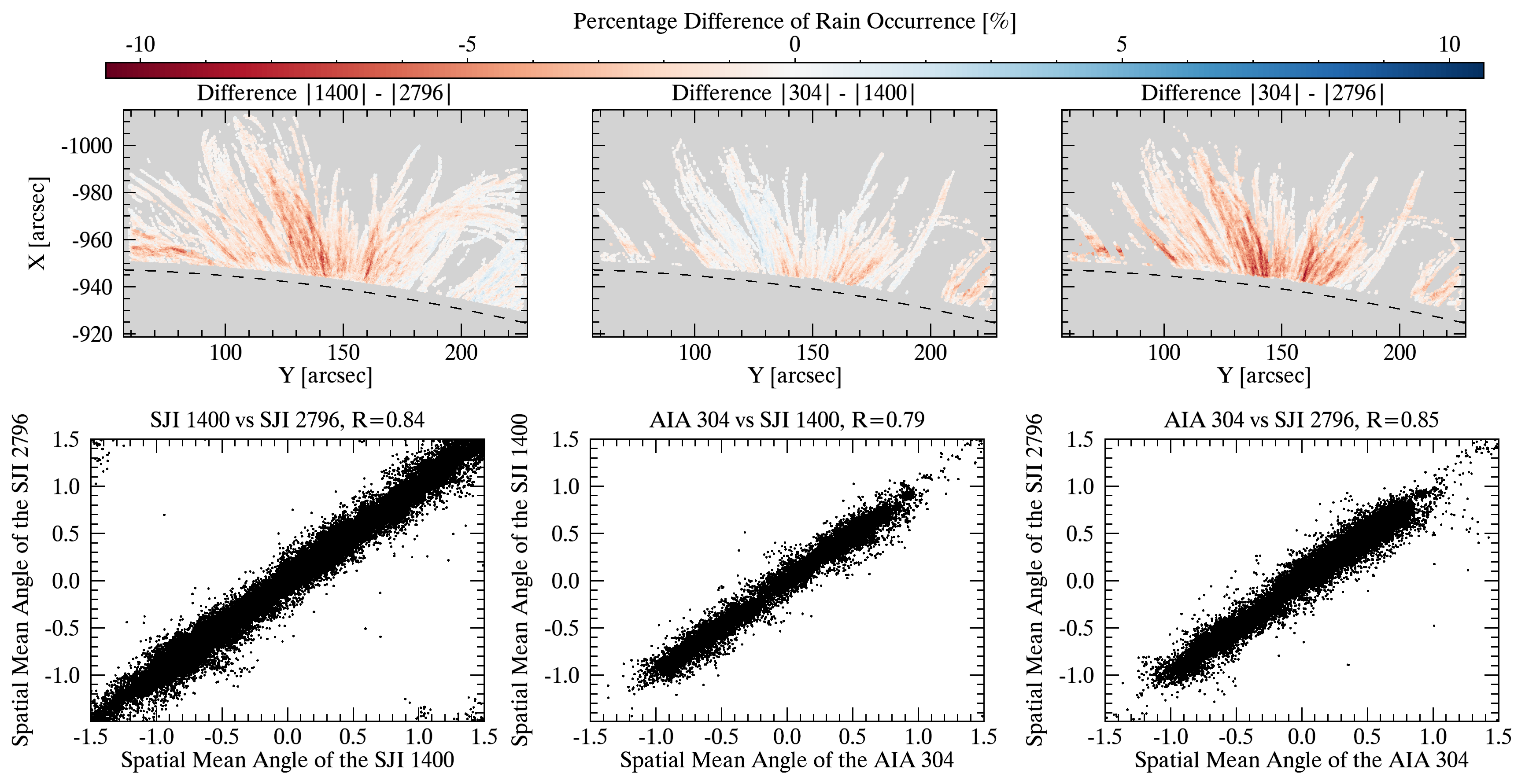}
\caption{Percentage difference maps of coronal rain occurrence (top) and the scatter plots (bottom) of the spatial mean angles given in Figure \ref{mean_angles}. We  calculate the total number of rain pixels over time in each x and y position for each channel, identifying common non-zero locations and taking the difference of absolute values between them. The resulting difference maps are divided by the total number of images to obtain a percentage difference. The dashed lines over the top plots indicate the solar limb. The Pearson correlation coefficient (R) for each pair is indicated in the bottom plots.} \label{dif_maps}
\end{figure*}

\subsection{Coronal rain detection}
To apply our semi-automatic coronal rain detection routine, we first remove the region around the solar limb to avoid its large brightness contrast. We also remove the spicular region above the limb, defining a minimum height (6 Mm) for the off-limb FOV in order to avoid most of the spicules and other low-lying, cool chromospheric features, which are typically seen at heights of 4-10 Mm above the solar surface \citep{Beckers1968SoPh....3..367B}. Then, we apply an automatic detection routine called the Rolling Hough Transform \citep[RHT;][]{schad2017} based on the Hough Transform \citep{hough1962method} (a technique for detecting lines, circles, and curvilinear structures) in order to detect and quantify the apparent motion of the raining material. Along the temporal axis, a running mean filter is applied to ensure that the rain clumps trace out some path for each individual time step. This local path is then used to determine a flow direction in the plane-of-the-sky (POS). For the SJI 1400~\AA, SJI 2796~\AA, and AIA 304~\AA~channels, we used a 10-step ($\approx$ 7.18 min), a 13-step ($\approx$ 6.99 min), and a 17-step ($\approx$ 3.4 min) running mean filter, respectively. Such values are chosen according to the observed speeds in coronal rain and thus the time it takes to see a significant change locally in the image. Subsequently, the code  applies a bidirectional difference filter along the temporal axis to assist in flow segmentation. For the IRIS channels, we use a 5-step ($\approx$ 3.6~min in SJI 1400~\AA~and $\approx$ 2.7 min in SJI 2796~\AA), and we use a 21-step filter ($\approx$ 4.2 minutes) for the AIA 304~\AA~to account for the faster cadence. After that, we use 29 pixels for the circular RHT kernel width ($D_w$) for all channels (i.e. the circular region with a diameter $D_w$ centered on each image pixel to be evaluated at one time). Choosing a wider kernel size leads to less detection of small rain clumps but reduces the background noise. Therefore, we choose this value according to the observed minimum length of the rain. Finally, the standard deviation accounting for the background noise $\sigma_{noise}$ is 2, 2.1, and 0.008 for the SJI~2796~\AA, SJI~1400~\AA, and AIA~304~\AA, respectively (the different noise values in 304~\AA~are due to the `Cool 304' pre-processing of images explained above). 

Using all these input parameters, the RHT routine produces four main outputs (each output has 3D dimensions with position $x, y$ and time $t$) providing spatial and temporal information for each. These are the mean axial direction, the peak of the RHT function, resultant length, and error in the mean axial direction. The mean axial direction ($\theta_{xy}$ for the spatial part and $\theta_{t}$ for the temporal part) shows the spatial and temporal directions of each detected pixel. To accurately define these mean directions, the RHT function ($H_{xy}(\theta)$ in the spatial part and $H_{t}(\theta)$ in the temporal part) is used to accumulate pixels to define a mean direction. For this purpose, an adaptive threshold is defined according to the peak of the RHT function ($max[H_{xy}(\theta)]$ and $max[H_{t}(\theta)]$), which is the second output of the RHT routine. Any values below this threshold do not contribute to the mean direction. The resultant length $(\Bar{R})$ is a measure of dispersion between 0 and 1. Values close to 0 or 1 indicate an ill-defined or well-defined mean axial direction, respectively. The last output is the error ($\epsilon_{xy}$ and $\epsilon_{t}$) in the mean axial direction. A detailed explanation of all these parameters can be found in \citet{schad2017}. In total, we detected 
4.33$\times$10$^6$, 2.11$\times$10$^6$, 3.25$\times$10$^6$ 
rain pixels in SJI~2796~\AA, SJI~1400~\AA, and AIA~304~\AA, respectively. These numbers are smaller than those detected in the same AR in \citet{Sahin2022ApJ...931L..27S} (6.5$\times$10$^7$, 4.9$\times$10$^6$, 6.33$\times$10$^6$ in the AIA 304~\AA, SJI 1400~\AA, and SJI 2796~\AA, respectively). This is because of the stricter conditions we have used in the RHT routine since we are interested in higher accuracy when calculating trajectories and projected speeds in our analysis. Accordingly, we have also required the following set of conditions.   We only include the pixels for which $\overline{R}_{xy}\geq0.8$, $max[H_{xy}(\theta)]\geq0.75$, $\overline{R}_{t}\geq0.8$, $max[H_{t}(\theta)]\geq0.75$ for all channels, and $\overline{|\theta}_{t}|\le84^\circ$ for the AIA 304~\AA~and $\overline{|\theta}_{t}|\le88^\circ$ for both SJI channels, since these values ensure relatively small errors for projected velocities. The remaining rain pixels for our analysis are 6.71$\times$10$^5$, 1.90$\times$10$^5$, and 2.83$\times$10$^5$ in the SJI~2796~\AA, SJI~1400~\AA, and AIA~304~\AA, respectively.

\subsection{EUV pulsation detection}

For the long-period pulsation analysis, we study 3-days data at a cadence of 5 minutes (between 2017.06.01 05:00 UT and 2017.06.04 05:15 UT) with AIA using a different version of the automatic detection algorithm used in \citet{Auchere2014A&A...563A...8A} and \citet{froment2015}, modified for off-limb observations \citep{froment2020}. In this routine, the power spectral density (PSD) is calculated for each time series associated with each pixel of the FOV, without specifically searching for pulsations in a particular region or structure. A 4×4 pixel binning is applied to increase the signal-to-noise ratio. Please see the above-mentioned papers for additional information regarding the detection algorithm.

\section{Results} \label{sec:result}

\subsection{Spatial and temporal mean angle and rain occurrence}

The average spatial ($\overline{\theta}_{xy}$) and temporal ($\overline{\theta}_{t}$) mean angle maps over an entire time sequence from the RHT analysis are shown in Figure \ref{mean_angles}. The spatial mean angle $\overline{\theta}_{xy}$ is related to the inclination of the rain in the POS, while the temporal mean angle $\overline{\theta}_{t}$ shows the dynamical change along the trajectory. All the colored pixels in these maps correspond to  coronal rain pixels. As seen, coronal rain occupies a wide POS area over the AR. It is worth noting that much more rain is seen to occupy the FOV (about a factor of 10 more rain pixels exist). However, as explained in the previous section, we choose to set strict conditions for the detection of coronal rain in order to more accurately capture the dynamics and trajectories.

The observed rain trajectories match well the expected large-scale magnetic field topology of the AR, given the sunspot at the centre of the FOV. It is worth noting that this is the first of its kind high-resolution coronal field line tracing on a large-scale with coronal rain. In the temporal mean angle maps (bottom maps in Figure \ref{mean_angles}), the green and red colours (negative values) represent downward motion which, as expected from the action of gravity on coronal rain,  is much more dominant. The total pixel number of downflow motions are $6\times10^5$,$1.6\times10^5$, and $1.9\times10^5$ for the SJI 2796~\AA, SJI 1400~\AA, and AIA 304~\AA, respectively, which represent 90.05\%, 83.44\%, and 89.11\% of the total detected rain pixels. Upward motions are also ubiquitous ($6.7\times10^4  (9.95\%)$, $3.1\times10^4 (16.56\%)$, and $3.1\times10^4 (10.89\%)$ for the SJI 2796~\AA, SJI 1400~\AA, and AIA 304~\AA, respectively), particularly near the sunspot, as seen in the bottom panel (blue). This is expected because gas pressure can also play an important role in coronal loops \citep{antolin2010}, a result supported by recent 2.5D numerical simulations \citep{Li2022ApJ...926..216L}. 

\begin{figure*}[ht]
\includegraphics[width=\textwidth]{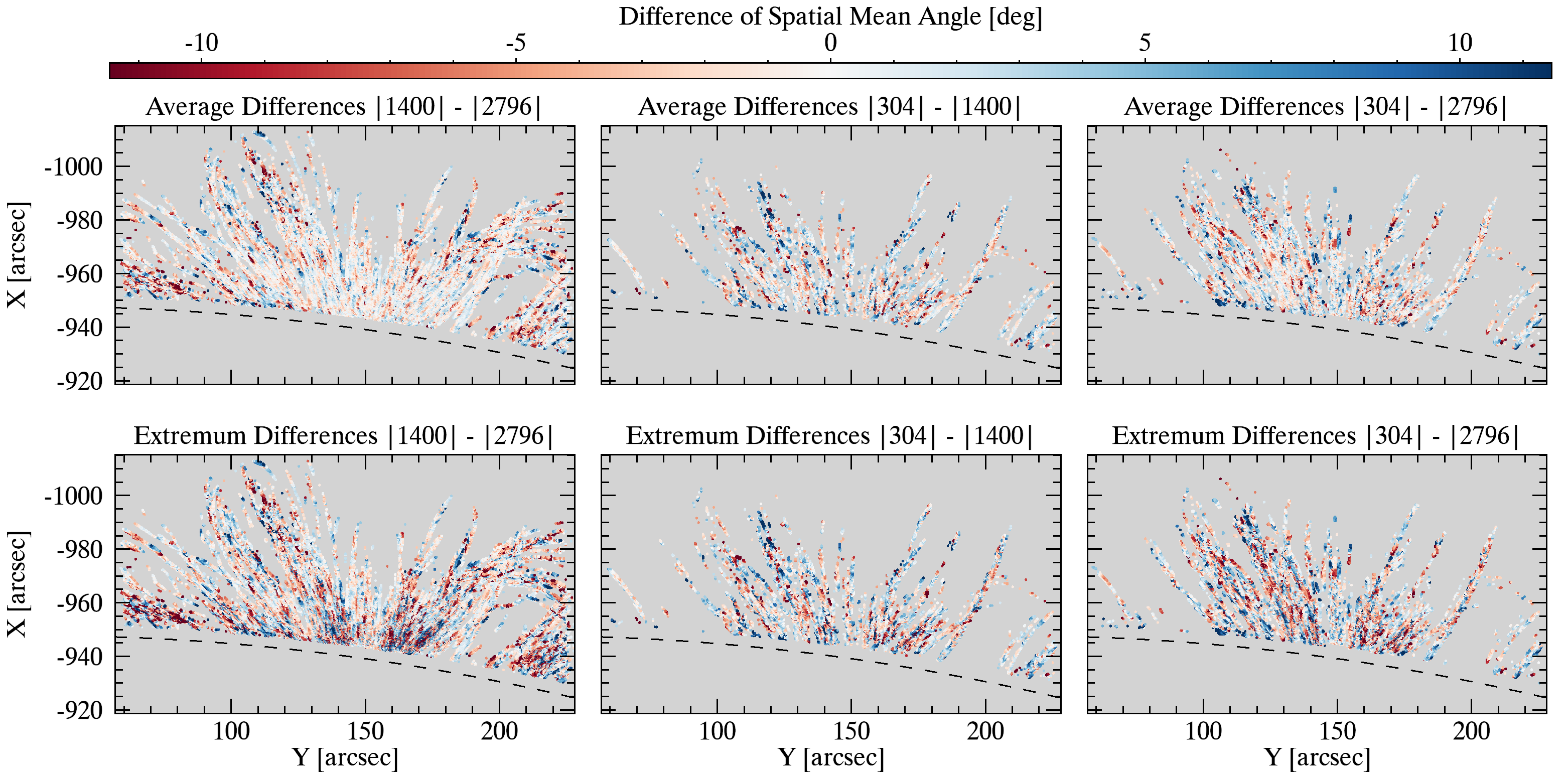}
\caption{Average (top) and Extremum (bottom) difference maps of absolute spatial mean angle values given in Figure \ref{mean_angles}. These values have been saturated by 11 degrees in order to see the spatial variation better. The dashed lines over the top plots indicate the solar limb. Please refer to the text for further details.}\label{dif_maps2}
\end{figure*}

In Figure \ref{dif_maps}, we investigate the difference in coronal rain occurrence across the channels. On the top and bottom panels, we show, respectively, maps of `percentage difference' in coronal rain occurrence and scatter plots of the spatial mean angle together with the Pearson correlation coefficient. Here, we use the absolute spatial mean angle values and an equal number of time steps across channels to obtain maps of the difference in coronal rain occurrence. Since the SJI~1400~\AA~has the lowest cadence, we select it as a reference channel and find the closest time to it on the SJI 2796~\AA~and the AIA 304~\AA. For each pair of cubes among this set of three, we first calculate the total number of rain pixels (defined as non-zero spatial mean angle values) over time in each $x$ and $y$ position and gather them in a new cube. Then, we find the locations common to both new cubes (one for each channel) where we have non-zero values. Finally, we take the difference between them and divide it by the total number of images to obtain a percentage difference in coronal rain occurrence. As shown by the scatter plots, coronal rain emission in all channels is highly correlated ($R=0.84$ between the SJI 1400~\AA~and 2796~\AA, 0.79 for the AIA 304~\AA~and SJI 1400~\AA, and 0.85 for the AIA 304~\AA~and the SJI 2796~\AA). This indicates that at a global level, coronal rain is multi-thermal, with chromospheric emission co-located on average with transition region emission. However, the normalised difference maps in the top panels reveal strong localised differences in all channels. From the top right and left panels, we see more rain in 2796~\AA~than in 1400~\AA~and 304~\AA~towards the footpoints. There is also more rain emission in the SJI~1400~\AA~than in the AIA~304~\AA~(top middle panel). These differences may also be due to intensity (opacity of the line) and contrast with the background, additional noise (low instrumental sensitivity, especially for the AIA 304~\AA), and the spatial resolution of the instrument.  

To investigate the differences in trajectories across the temperature range, we show in Figure \ref{dif_maps2}, the average (top) and extremum (bottom) difference maps of absolute spatial mean angle values. Similar to Figure \ref{dif_maps}, we used an equal number of images for the channels. Here, we first find the locations common to both channels where we have non-zero values, and then we take the difference between them. Then, for a given time t we take a time length around t equal to the mean running filter (chosen for the RHT routine) so that significant change is seen locally. We calculate the average and extremum (maximum in absolute values, retaining the sign of the difference) of the differences over this time range (over non-zero instances) between the two channel pairs, and store the average and extremum in two new cubes. We repeat this process for all spatial pixels, and we then calculate the average and extremum over time for these two cubes in the top and bottom panels in Figure \ref{dif_maps2}, respectively, over all non-zero time instances. The differences across channels are below 10 degrees on average, while the extrema hover around 10 degrees with isolated cases of up to 20 degrees (the figure is saturated at 11 degrees to best observe the trends). The $|1400|-|2796|$ panels reveal that differences occur on average in regions characterised by inclined structures or at the apexes of loops. We also observe increased variations close to the sunspot or solar surface. On the other hand, the other panels ($|304|-|1400|$ and $|304|-|2796|$) show a more spatially uniform distribution of the differences. It is likely that the reduced accuracy of the RHT method in the near loop apexes, attributable to the slower speeds of clumps in these regions, contributes to the observed differences in these locations. However, for structures characterized by higher speeds, such as those near the sunspot and inclined structures, the observed differences are more likely to be genuine. We can think of two possibilities. They may be the result of the superposition of loops along the LOS, particularly for the SJI 1400~\AA, which is usually optically thin. Similarly, the narrow width of the rain may reduce the optical thickness of the SJI 2796~\AA. Another possibility is multi-thermal plasma moving differently in the same pixel elements, which would lead to shear motions. 


\subsection{Rain Clumps Dynamics} \label{subsec:dynamics}

\begin{figure*}[ht]
\includegraphics[width=\textwidth]{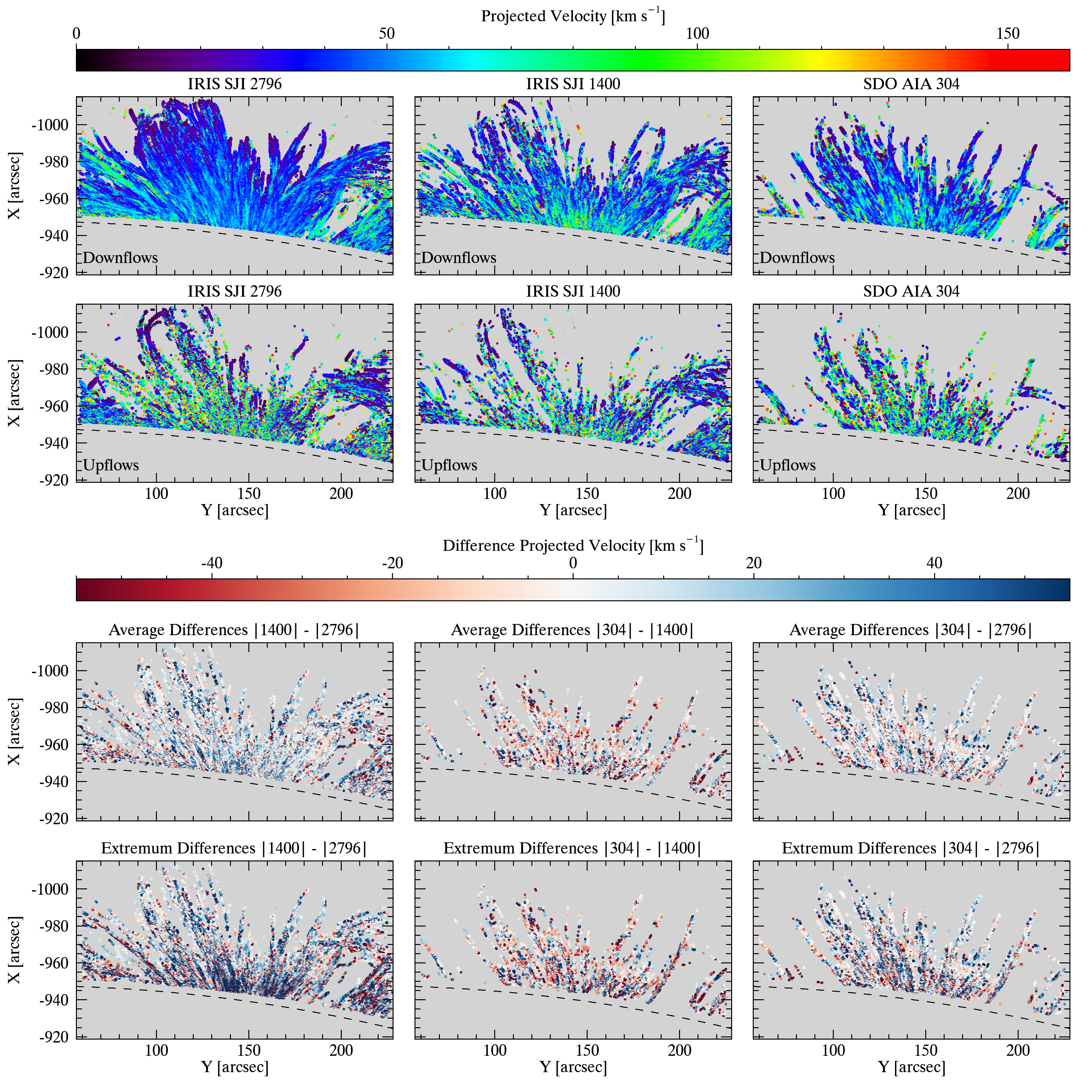}
\caption{Average downflow (top row) and upflow (second row from top) projected velocities in SJI~2796~\AA\ (left), SJI~1400~\AA\ (middle), and AIA~304~\AA\ (right). The average and extremum differences of the projected velocities saturated by 55 km $s^{-1}$ (panels in the last two bottom rows). The dashed lines over the plots indicate the solar limb. Please see the text for more details.} \label{rain_dynamics}
\end{figure*}

In this section, we show the dynamical results obtained through the RHT analysis. In Figure~\ref{rain_dynamics} (top panels), we show projected velocity maps resulting from the individual measurements of the rain clumps. The velocity along each curvilinear path obtained from the RHT temporal mean angle $({\overline{\theta}_{t}})$ is \citep{schad2017}:
\begin{equation}
\label{vparalel}
v_{||}=\tan{\overline{\theta}_{t}}(\frac{\delta x}{\delta t}).
\end{equation}
Here $\delta x$ is the spatial sampling on the date, which is 244.72~km for the SJI channels and 244.75~km for the AIA channel, and the $\delta t$ is the average cadence, which is 43.1~s, 32.3~s, and 12~s for the SJI 1400~\AA, 2796~\AA, and AIA 304~\AA, respectively. 

The projected velocity is also related to the spatial mean angle $\overline{\theta}_{xy}$ through the horizontal (Equation \ref{vhorz}) and vertical velocities (Equation \ref{vvert}):

\begin{equation}
\label{vhorz}
v_{x}= v_{||} \cos \overline{\theta}_{xy}
\end{equation}

\begin{equation}
\label{vvert}
v_{y}= v_{||} \sin \overline{\theta}_{xy}
\end{equation}

The projected velocity can also be decomposed in terms of the tangential (Equation \ref{tan}) and radial (Equation \ref{rad}) velocity components:

\begin{equation}
\label{tan}
v_{tan}=v_{x}(-\cos{(\alpha))}+v_{y}(-\sin{(\alpha))} \\
\end{equation}

\begin{equation}
\label{rad}
v_{rad}=v_{x}(-\sin{(\alpha)})+v_{y}\cos{(\alpha)}, \\
\end{equation}
where $\alpha$ denotes the angle between the radial and horizontal directions in radians. The projected velocity is then:
\begin{equation}
\label{prov}
 v_p=|v_{||}|=\sqrt{v_{tan}^2 + v_{rad}^2}
\end{equation}



The average projected velocity of SJI 2796~\AA\, SJI 1400~\AA\, and AIA 304~\AA\ are shown in Figure \ref{rain_dynamics} with respect to rain direction (downflows and upflows). As seen in these maps (first two rows from the top), most of the coronal rain clumps accelerate during the fall. Correspondingly, higher velocities can be clearly seen in the centre of the active region (near the sunspot) in all three channels for downflow motions. Interestingly, this is also the case for the upflows, which suggest an increasingly larger role of gas pressure at lower heights (see Discussion section). In the last two bottom panels in the same Figure~\ref{rain_dynamics}, we show the differences between projected velocities across the channels by following the same method as for Figure~\ref{dif_maps2}. We note a salt-and-pepper distribution of highly localised red and blue values, which may again correspond to times for which a rain pixel is detected in one channel but not the other channel. However, some interesting large-scale differences emerge. For instance, in the ${|1400|-|2796|}$  panel, more blue pixels are seen near the sunspot, suggesting higher velocities ($\approx20$~km~s$^{-1}$ higher) in 1400~\AA\ than in 2796~\AA. The $|304|-|2796|$ panel shows larger patches of the loops that are consistent with one colour, and higher speeds seem to be found in 304~\AA\ near the sunspot, consistent with the previous result. The $|304|-|1400|$ panel shows more red than blue, suggesting higher 1400 speeds.
This result suggests the presence of multi-thermal shear flows and/or density (and temperature) dependent speeds. However, the drag effect, which is dependent on density predicts the opposite result, with higher speeds for denser (and therefore cooler) rain.


The number of snapshots in the datasets (SJI 2796~\AA, SJI 1400~\AA, and AIA 304~\AA) is different between the channels. Also, we have different exposure times, which change the characteristics of the rain (particularly the length of rain clumps). Therefore, we calculated the rain area per second in each channel using the number of snapshots, pixel size, and exposure times (the unit is $arcsec^2 s^{-1}$) as given in Equation \ref{rain_area_sec}.

\begin{equation}
\label{rain_area_sec}
\text{Rain area per second}=(N_{pixels}/n_{t})\times (dx dy/D_{t}) \\
\end{equation} 
where $N_{pixels}$ is the total number of pixels in a given channel, $n_{t}$ is the number of snapshots, $dxdy$ is the pixel size, and $D_t$ is the exposure time. 

\begin{figure*}[ht]
\includegraphics[width=\textwidth]{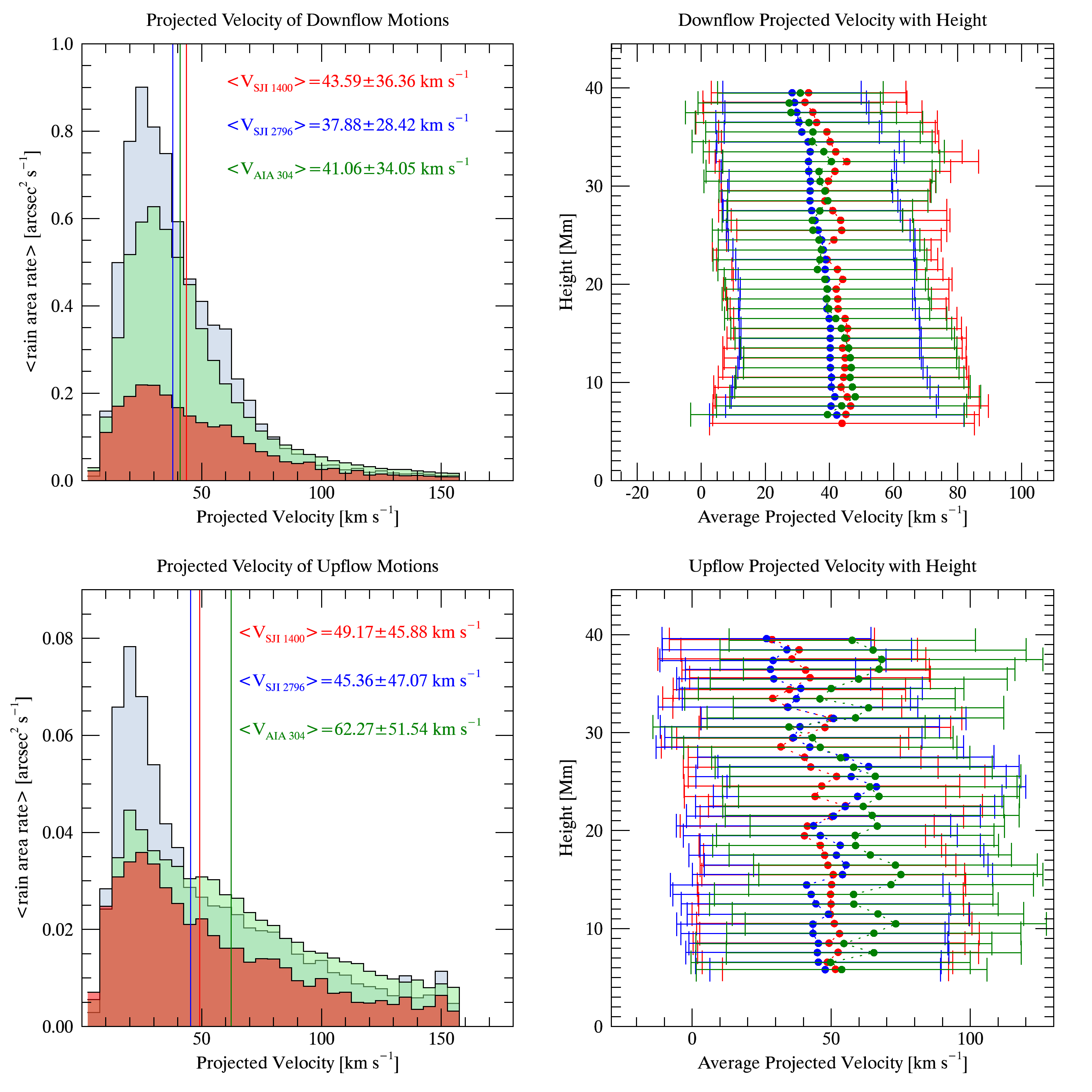}
\caption{Top: 1D Histogram distribution of the projected velocities for downflows (left) and their height distribution (right) in the SJI~1400~\AA\ (red), SJI~2796~\AA\ (blue), and AIA~304~\AA\ (green), with the corresponding median and standard deviation in the inner caption. Bottom: Same as above but for the upflows.} \label{rain_dynamics2}
\end{figure*}

The histogram of projected velocities in Figure~\ref{rain_dynamics2} on the left panels shows a broad distribution of downflow and upflow projected velocities, with peaks below 50~km $s^{-1}$ and long tails to high-velocity downflows (up to 200~km $s^{-1}$). The overall shape of the distributions is very similar across the channels, and consequently, the median velocities are very similar as well, with 43.59$\pm36.36$~km $s^{-1}$, 37.88$\pm28.42$~km $s^{-1}$, and 41.06$\pm34.05$~km $s^{-1}$ in SJI~1400~\AA, SJI~2796~\AA, and AIA~304~\AA, respectively for the downflow projected velocity. On the other hand, these are 49.17$\pm45.88$~km $s^{-1}$, 45.36$\pm47.07$~km $s^{-1}$, and 62.27$\pm51.54$~km $s^{-1}$ for the upflow projected velocity in the SJI~1400~\AA, SJI~2796~\AA, and AIA~304~\AA, respectively. The variation of the median projected velocities for both downflow and upflow motions with projected height is shown on the right panels in Figure \ref{rain_dynamics2} for SJI 1400~\AA~(red), SJI 2796~\AA~(blue) and AIA 304~\AA~ (green), where zero height corresponds to the solar limb, also shown in Figure \ref{rain_dynamics} with dashed lines on the panels. The median projected velocities for the downflow motions are quite similar at all heights, particularly between SJI 2796~\AA\ (blue) and AIA 304~\AA\ (green), SJI 1400~\AA\ has slightly larger velocity values at higher heights ($\geq 18~Mm$). This is in agreement with the results from Figure~\ref{rain_dynamics}. This plot also indicates that the rain experiences small acceleration at higher heights (close to loop apexes) and then continues downward with an almost constant velocity, which is expected from gas pressure forces \citep{Oliver2014}. Relative to the downward motion, the upflow projected velocities show much more variation with height, suggestive of strong gas pressure changes associated with the rain’s fall. The high speeds found in the histogram of Figure~\ref{rain_dynamics2} therefore correspond to highly localised measurements which can be due to apparent effects or errors in the RHT routine and not to large-scale bulk flows due to gravity. 

\begin{figure*}[ht]
\includegraphics[width=\textwidth]{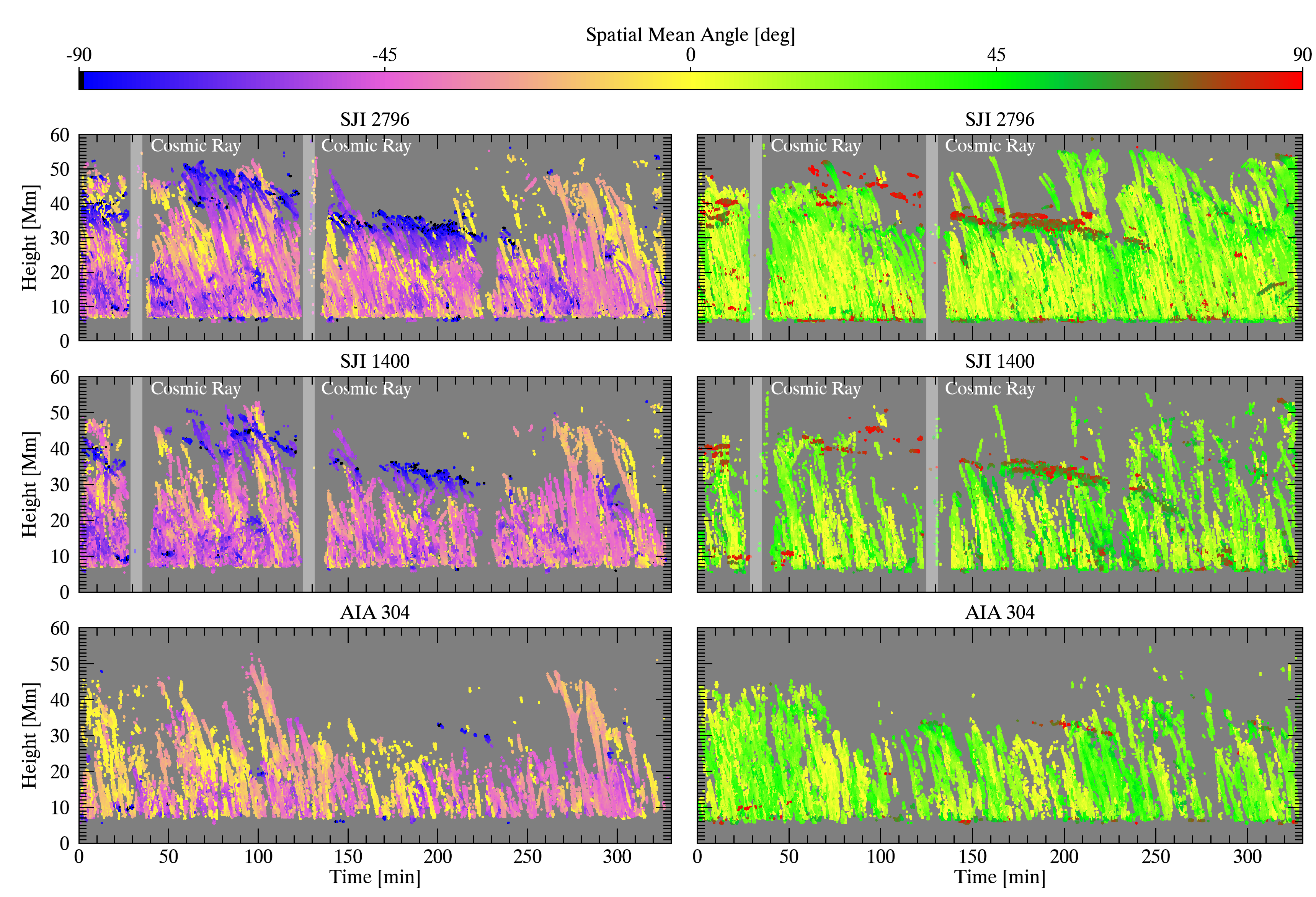}
\caption{Space-time diagram of spatial mean angle for the SJI 2796~\AA\ (top), SJI 1400~\AA\ (middle), and AIA 304~\AA\ (bottom). The grey shaded area on SJI maps corresponds to times of cosmic rays, in which the detection of coronal rain is affected. \label{timdis}}
\end{figure*}

In order to study the time evolution of the coronal rain dynamics, we plot in Figure \ref{timdis} the time-distance diagrams of the spatial mean angles seen in the SJI~1400~\AA\ (top), the SJI~2796~\AA\ (middle), and the AIA~304~\AA\ (bottom). Here, we overlay all time-distance maps for each radial axis and present the positive (on the right plots) and negative (on the left plots) spatial mean angle inclinations separately. Coronal rain (clumps and showers) is continuously observed during the entire time duration of the observations and across the full radial extent of the IRIS data. Although there are many overlapping rain trajectories over time, Figure \ref{timdis} reveals that the spatial mean angle changes with height and in time (for individual trajectories), suggesting a change in the rain's projected velocity as it falls. This change suggests an acceleration, but it is likely an apparent acceleration caused by the small angle between the LOS and the plane of the loop (which leads to slow speeds for rain near the loop apex). On the other hand, Figure \ref{rain_dynamics2} shows a linear increase in velocity with decreasing height, indicating a small constant acceleration. This result can be explained by gas pressure restructuring produced by the condensation formation \citep{Oliver2014, Martinez-Gomez2020}. The change in the slope could actually not correspond to acceleration but a change in the angle between the LOS and the vector tangent to the flow. If the LOS makes a small angle with the loop plane, the angle between the LOS and the vector tangent to the flow would be close to zero at the apex of the loop, which makes the projected velocity very small, while towards to loop leg, the angle would increase toward 90 degrees, which makes the projected velocity close to the total velocity.

\begin{figure*}[ht!]
\plotone{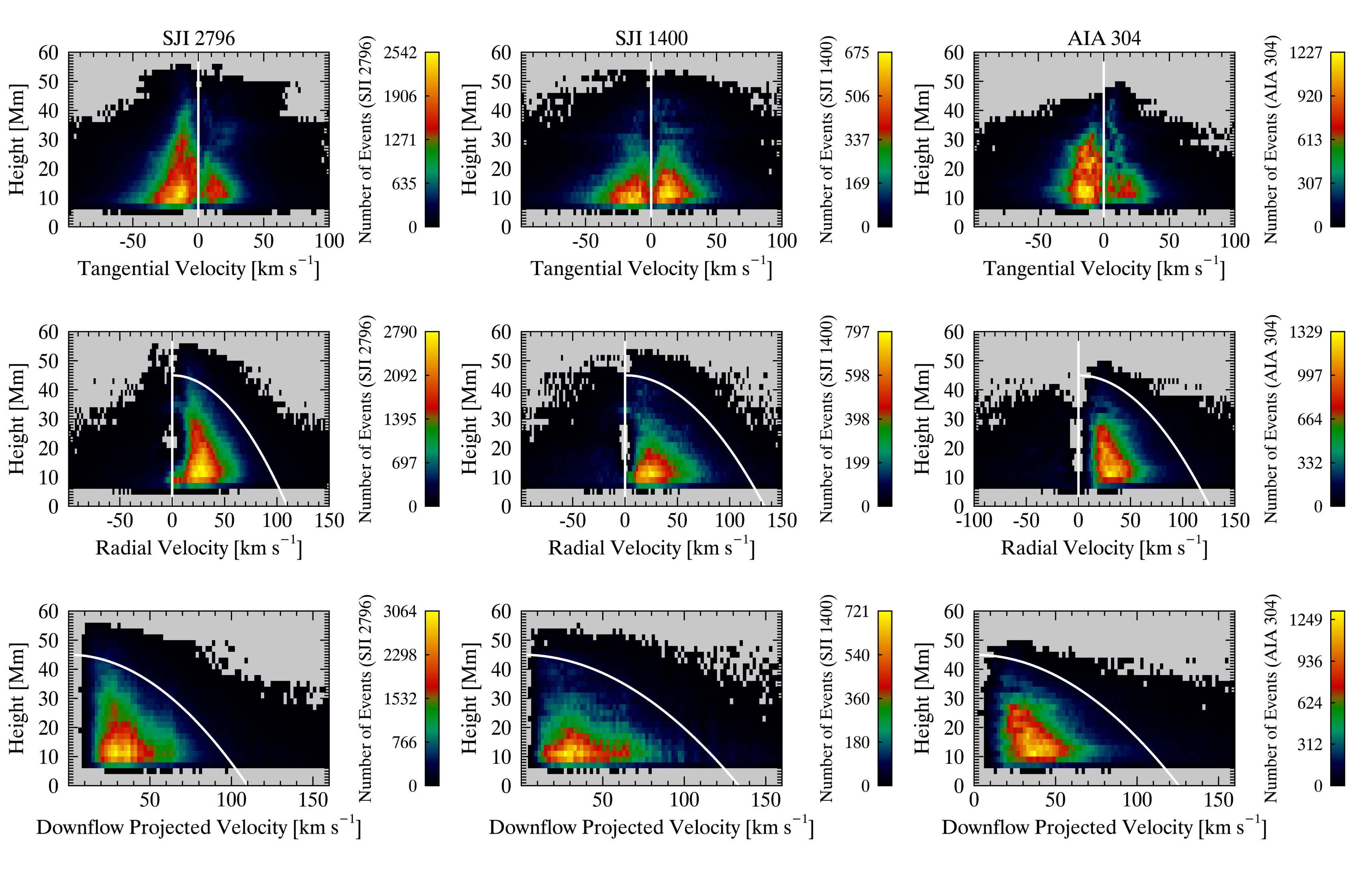}
\caption{Two-dimensional probability distribution functions (PDFs) of the tangential, radial, and downflow projected velocities. The white parabola denotes the free-fall velocity change with height, taking as starting height the observed average. \label{2dpdf}}
\end{figure*}

In Figure \ref{2dpdf}, we show the 2D probability distribution function (PDF) of the tangential, radial, and downflow projected velocities relative to the height of the rain material. As seen in Figure \ref{mean_angles}, rain clumps appear on both sides of the AR in a roughly equal manner and this is seen as a bi-modal distribution in the tangential velocity maps (top panels). In the middle and bottom panels, we can see that the downward velocities are consistently lower than the free-fall velocity limit (taking an average starting position from the rest of 45~Mm), in agreement with previous results \citep{schrijver2001, DeGroof2004, antolin2010}. 

\subsection{Morphology\label{subsec:morph}}

\begin{figure*}[ht]
\includegraphics[width=\textwidth]{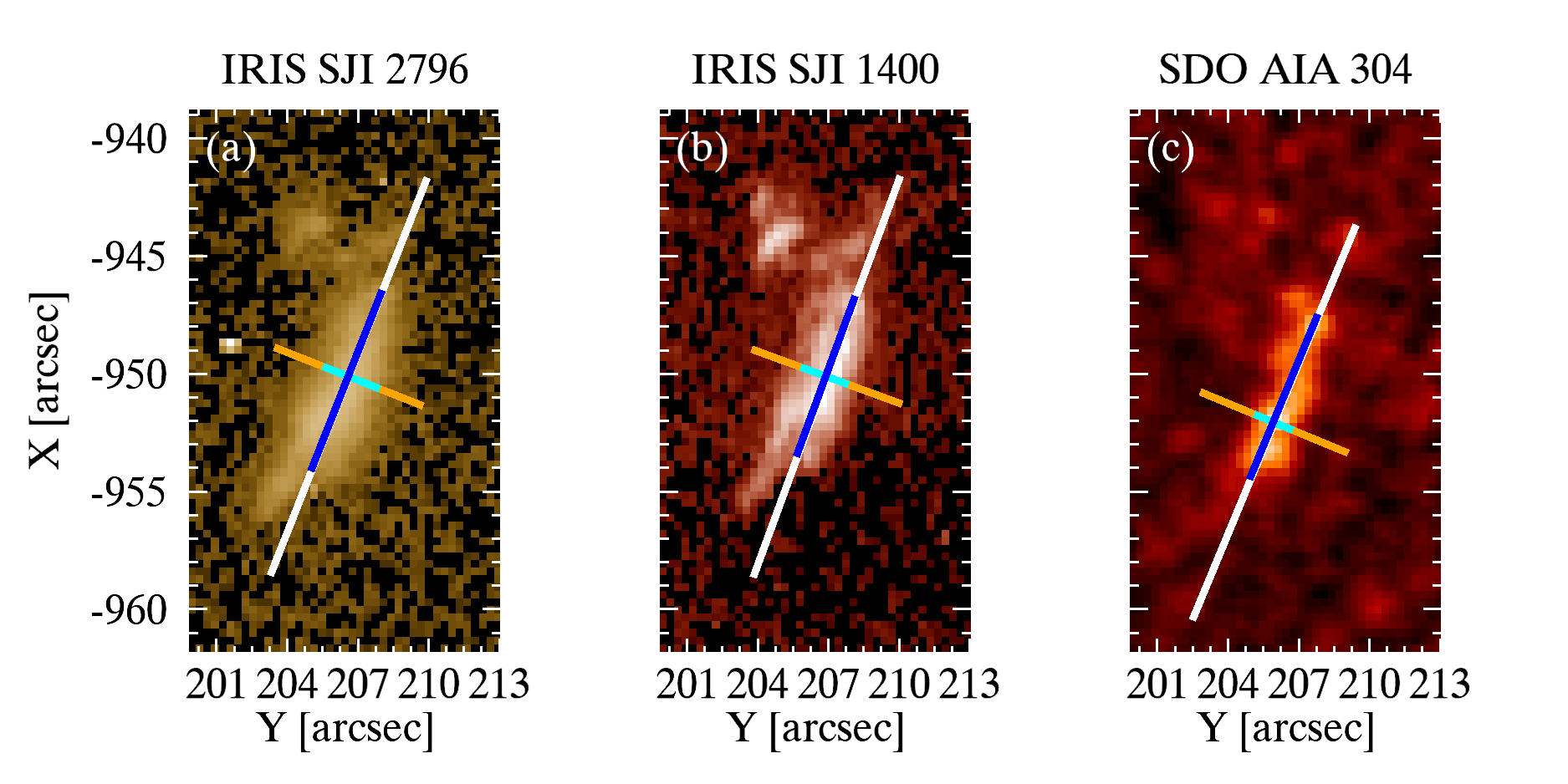} \\
\includegraphics[width=\textwidth]{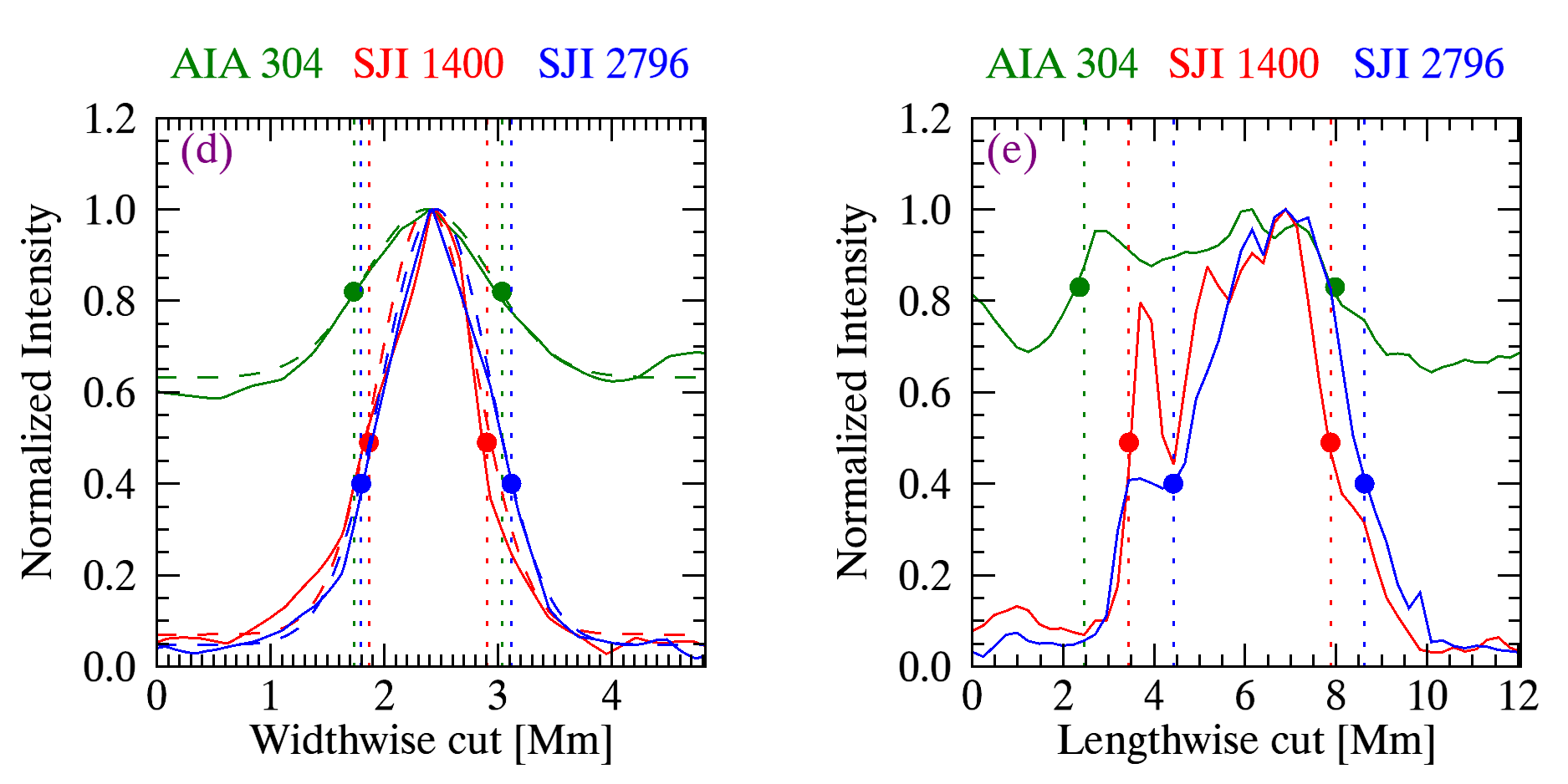}
\caption{Method for calculating the rain clump's width (d) and length (e) in SJI~2796~\AA\ (a), SJI~1400~\AA\ (b), and AIA~304~\AA\ (c). The orange line in each top panel shows a cut crossing the rain clump at a particular pixel, perpendicular to the clump trajectory, which is used to calculate the clump's width from a Gaussian fit to the intensity profile across that line (FWHM corresponds to the cyan part). The white lines denote the cuts along the clump that are used to calculate its length (corresponding to the blue part). The intensity profiles of the clump along the orange (d) and white (e) cuts are shown in the bottom panels, together with the resulting extrema positions indicating the width and length (dashed lines in each panel with green, red and blue corresponding to AIA~304~\AA, SJI~1400~\AA\ and SJI~2796~\AA, respectively). Filled circles correspond to the intensity positions at FWHM. See the main text for more detail.}  \label{noise}
\end{figure*}
We now turn to the statistical determination of the size of individual rain clumps. The width of the clumps is calculated in the following way. Given a clump pixel at a particular time and location, we first obtain the spatial mean angle information. We then draw a perpendicular line to the trajectory of the clump at that pixel and fit a Gaussian to the intensity profile interpolated over the perpendicular line (see the orange line in Figure \ref{noise}). From the fit, we obtain the full width at half maximum (FWHM), which we take as the width for the rain clump at that pixel. 


We also set a condition on the intensity of the profile to avoid background noise, and we set an intensity threshold below which we neglect the intensity profiles. The thresholds for each channel are calculated by selecting a small region without any visible structure and calculating the average intensity over the entire observational sequence. 
As we mentioned in Section \ref{sec:method}, the noise level across the image and during one raster is non-uniform because of the scattered light and imperfect flat fielding. Although this non-uniformity is corrected to some degree in our data calibration (see section~\ref{sec:method}) it is not entirely removed. We, therefore, take into account a spatially and temporally non-uniform secondary noise threshold that is used for the calculation of the length. Given a rain pixel detected by the RHT, the width calculation procedure explained above includes a Gaussian fitting whose FWHM sets the width. As seen in Figure~\ref{noise}, this procedure also sets an intensity threshold, the intensity at half maximum (filled circles corresponding to the positions of the dotted lines in panel d), which we use as the non-uniform intensity threshold for the length calculation (intensity values corresponding to the positions of the dotted lines in panel e).


To calculate the clump length, given a pixel location, we first take the centre point of the Gaussian fit used to calculate the clump's width at that location. Using the spatial mean angle at this centre point, we draw a path along the same trajectory and determine the locations along this path at which the intensity of the clump falls below the non-uniform noise threshold explained above. These naturally lead to two points above and below the centre point, which define the extrema for the length calculation. The average of all the length calculations for all pixels of a given clump would provide an accurate measurement of the clump's length. An example of this method is shown in Figure \ref{noise}.

\begin{figure*}[ht]
\includegraphics[width=\textwidth]{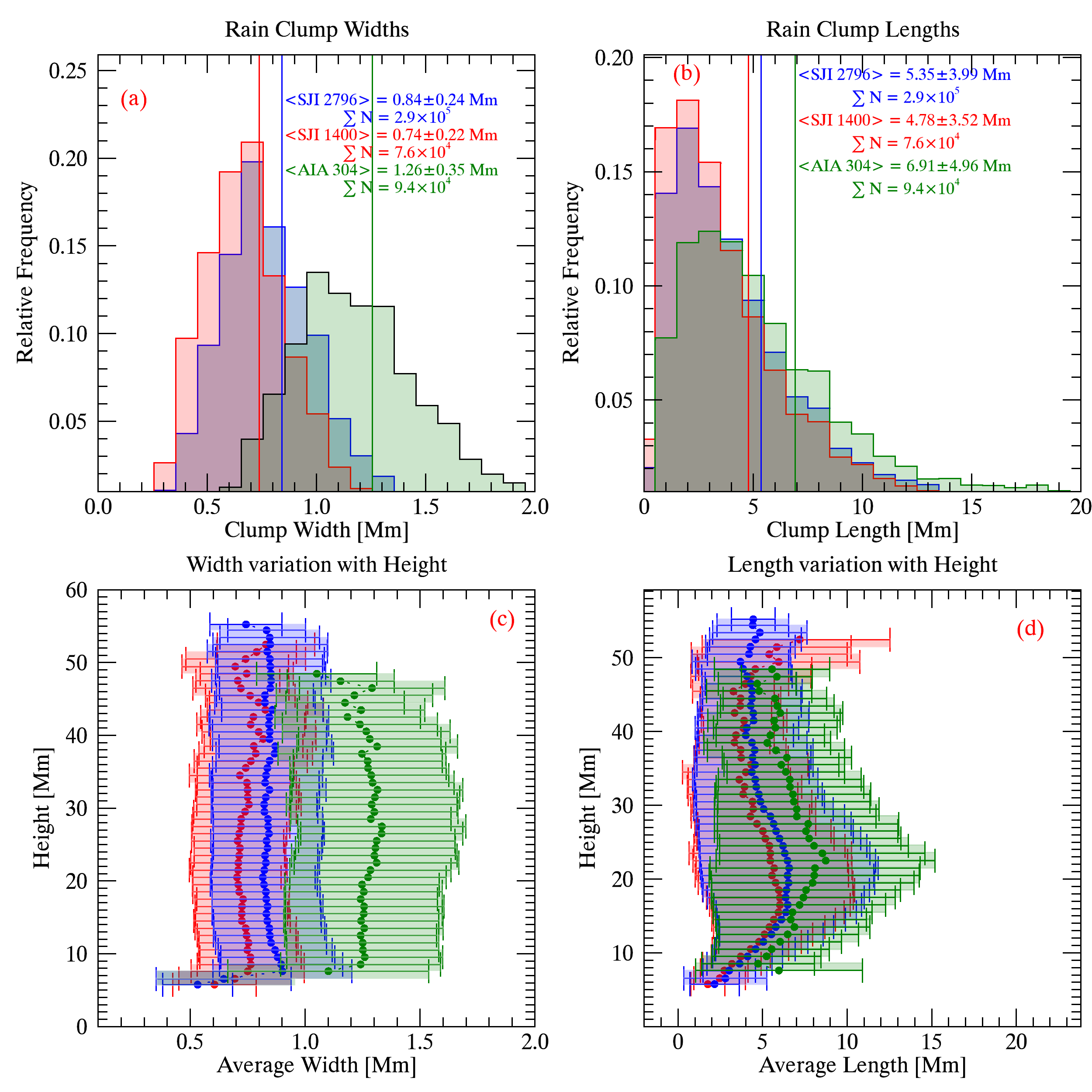}
\caption{Histograms of rain clump widths (a), lengths (b), and their variation with height (c,d), respectively. Blue, red and green denote, respectively, SJI~2796~\AA, SJI~1400~\AA, and AIA~304~\AA\ (see legend). In the bottom panels, circles and error bars correspond to the average width (c) and length (d), and their average standard deviation at each height bin.  \label{morphology}}
\end{figure*}

The histogram plots in Figure \ref{morphology}{a,b} show the widths and lengths distributions of all the pixels of the traced coronal rain clumps in all three channels. The distributions are similar in shape across all channels and are slightly asymmetric with a small tail at higher values. The average widths are found to be 0.84$\pm$0.24~Mm for SJI~2796~\AA, 0.74$\pm$0.22~Mm for SJI~1400~\AA, and 1.26$\pm$0.35~Mm for the AIA~304~\AA. The lengths, on the other hand, are found to be 5.35$\pm$3.99~Mm, 4.78$\pm$3.52~Mm, and 6.91$\pm$4.96~Mm for the same channel ordering. The widths between SJI~1400~\AA\ and SJI~2796~\AA\ are, therefore, similar, but the widths found in AIA~304~\AA\ are significantly larger. This may be explained by the lower spatial resolution in AIA~304~\AA\ and the fact that the rain widths are only a few times higher than the instrument's resolution ($\approx 0.87$ Mm for the AIA and $\approx 0.48$ Mm for the SJI channels). A similar result but to a lower degree is seen for the lengths, which are an order of magnitude larger than the widths. The bottom plots (c,d) show the width and length variations with height above the solar surface ($z=0$). The widths are fairly constant as the rain falls; however, the lengths vary significantly, becoming shorter as the rain falls (passing from $\approx$15 Mm to 6 Mm). The reason for the shortening length of the clumps is probably due to the fact that they cross below the limb, and we are, therefore, unable to track them further. Another reason may be that the rain also elongates (i.e. becomes less clumpy), which is supported by the increasing trend in Figure \ref{morphology}{d} between 20~Mm and 40~Mm, in which case the temporal segmentation approach is less accurate.



\subsection{Long and short term periodicity \label{subsec:period}}

\begin{figure*}[ht]
\includegraphics[width=\textwidth]{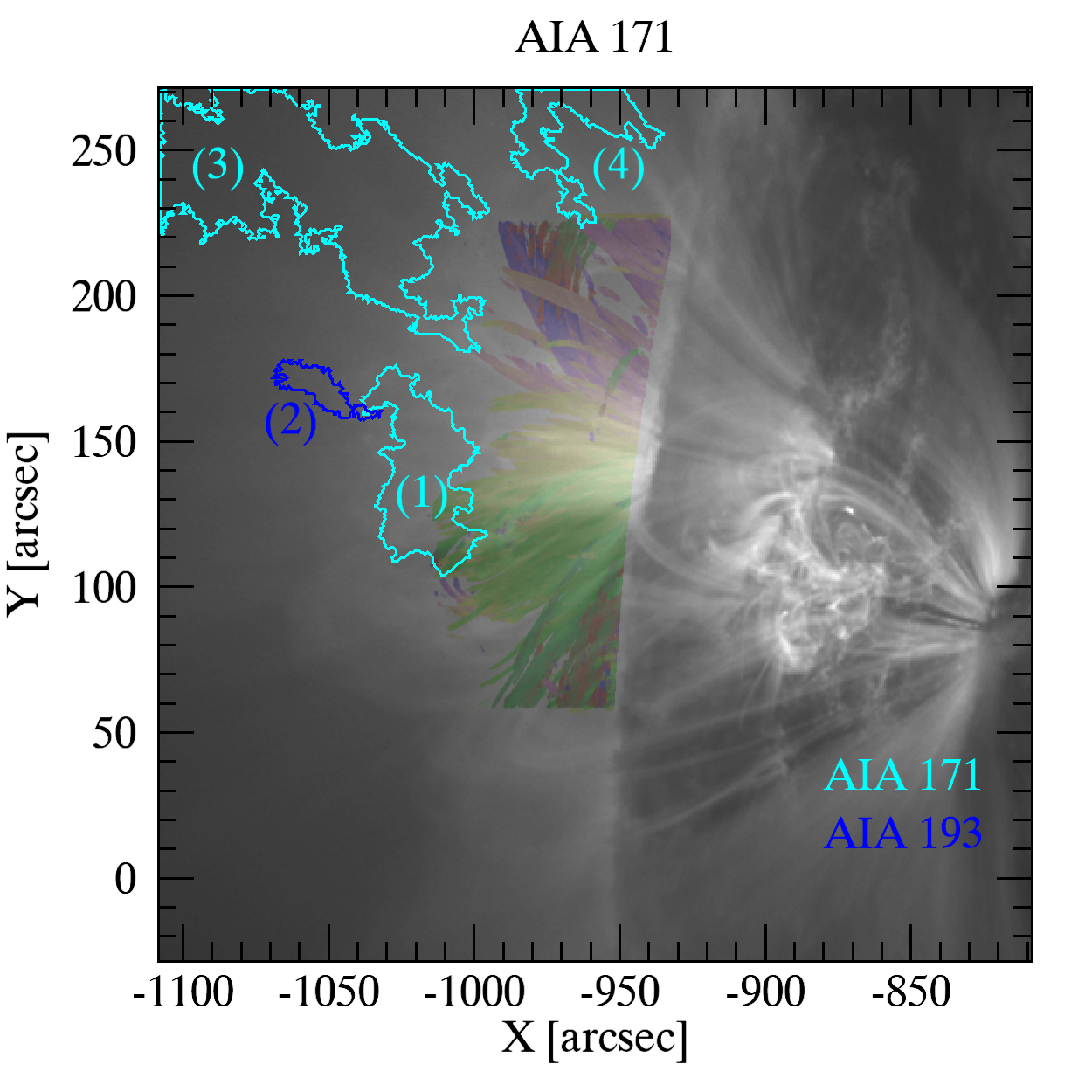}
\caption{Location over the active region of the long-period EUV intensity pulsations in AIA 171~\AA\ (cyan contours) and AIA 193~\AA\ (dark blue contours) with normalised Fourier power above 10~sigma. Each contour is labelled with the corresponding dominating frequency: 75.97 $\mu Hz$/3.65~hr (1), 65.28 $\mu Hz$/4.25~hr (2), 38.40 $\mu Hz$/7.23~hr (3), and 49.92 $\mu Hz$/5.56~hr (4). We overlay the spatial mean angle map obtained with the RHT from the IRIS~2796 images. \label{pulsation_rht}}
\end{figure*}

We now check for the existence of long and short-period intensity pulsations using the 3-days data at a cadence of 5 minutes (between 2017.06.01 05:00 UT and 2017.06.04 05:15 UT). In Figure~\ref{pulsation_rht}, we show the contours of some of the detected pulsations regions in AIA~171~\AA\ (cyan) and AIA~193~\AA\ (dark blue) over an AIA~171 image of the AR with the overlaid 2796 spatial mean angle map from coronal rain given in Figure~\ref{mean_angles}. These contours correspond to regions with more than 10$\sigma$ of normalized power. The pulsations are mostly found at the top or apex of the coronal loop structures, probably due to the reduced superposition of bright coronal structures at higher heights.  

\begin{figure*}[ht]
\includegraphics[width=\textwidth]{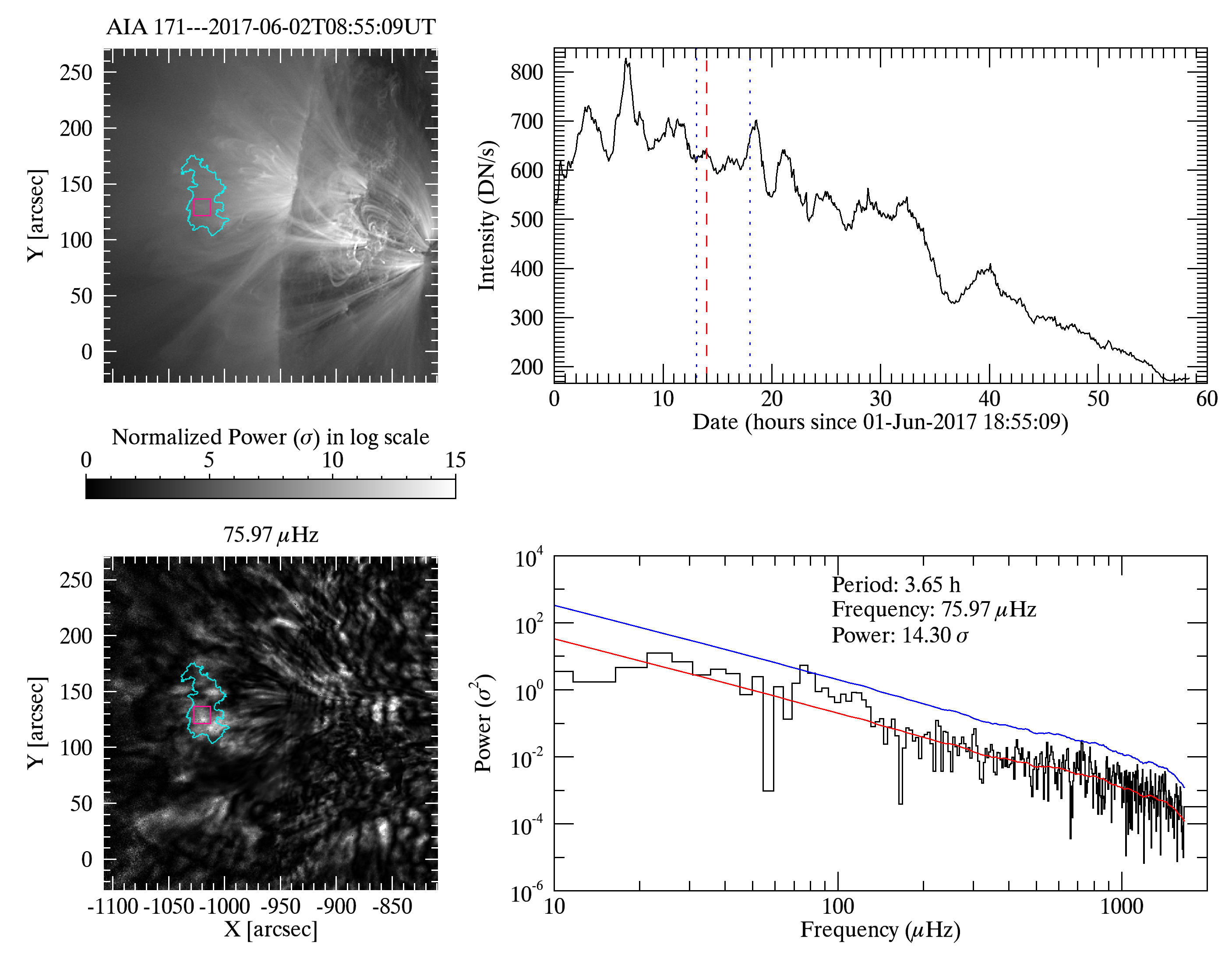}
\caption{Top Left: AIA 171~\AA\ map with one of the detected EUV intensity pulsation regions (cyan contour). The pink square is the sub-region for the detailed analysis. Top Right: AIA~171~\AA\ intensity evolution over the observational period summed over the sub-region. The dashed vertical line shows the time of the right map, while the vertical blue dotted lines indicate the observation time window of the IRIS instrument (i.e. 5.45 hr). Bottom Left: The normalized power map of the AIA~171~\AA\ in logarithmic scale at 75.97~$\mu$Hz (3.65~hr). Bottom right: Fourier power spectrum of the AIA~171~\AA\ lightcurve. The solid red curve shows the estimation of the average local power, and the blue curve is the 10~$\sigma$ detection threshold. Accompanying movies are also available. The movie has the same layout as the figure and shows the evolution of the active region in AIA 171~\AA~and the corresponding normalised power map at a frequency of 75.97 $\mu$Hz over the 2.4-days period around the IRIS observation time. In the movie, we can see the AR rotating into the disc starting from 1 June 2017, 18:55:09 UT to 3 June 2017, 05:15:09 UT. The red dashed vertical line in the top right panel shows the time of the top right panel.\label{fourier_puls}}
\end{figure*}

To further check the existence of periodicity, we repeated the Fourier power calculation for manually selected, smaller areas inside these regions and found periods between 3 and 9 hours. In Figure~\ref{fourier_puls} we show one example of such analysis, with the pink square in the AIA~171 map of the selected sub-region that we focused on for the detailed analysis. The AIA~171~\AA\ lightcurve on the top-right panel shows the time evolution over the 2.4-day period of the intensity variation summed over the pink square. The bottom panels show the corresponding Fourier spectra (right) and the map of normalized power (left) for the AIA~171~\AA\ passband in logarithmic scale at the frequency of the highest power peak 75.97$\mu$Hz (~3.65 hours) with the same field of view as in the top-left panel. The solid red curve in the Fourier spectrum is the estimate of the average local power (and therefore, a representative of the noise level), and the blue curve is the 10$\sigma$ detection threshold. In the 131~\AA\ and 193~\AA\ channels, we find power peaks at the frequency of 76.8$\mu$Hz, with a normalised power of 11.33$\sigma$ and 8.77$\sigma$, respectively. For the remaining channels (94~\AA, 211~\AA, and 335~\AA), there is no significant power. The power in 171~\AA\ is the highest, with 14.30$\sigma$. Even though it has been not shown in this study, we have also checked this periodicity with a wavelet analysis that shows a 95\% confidence level for a significant power distribution (13.94$\sigma$) between 18:55:09 and 3:15:09 time range.

\begin{figure*}[ht]
\includegraphics[width=\textwidth]{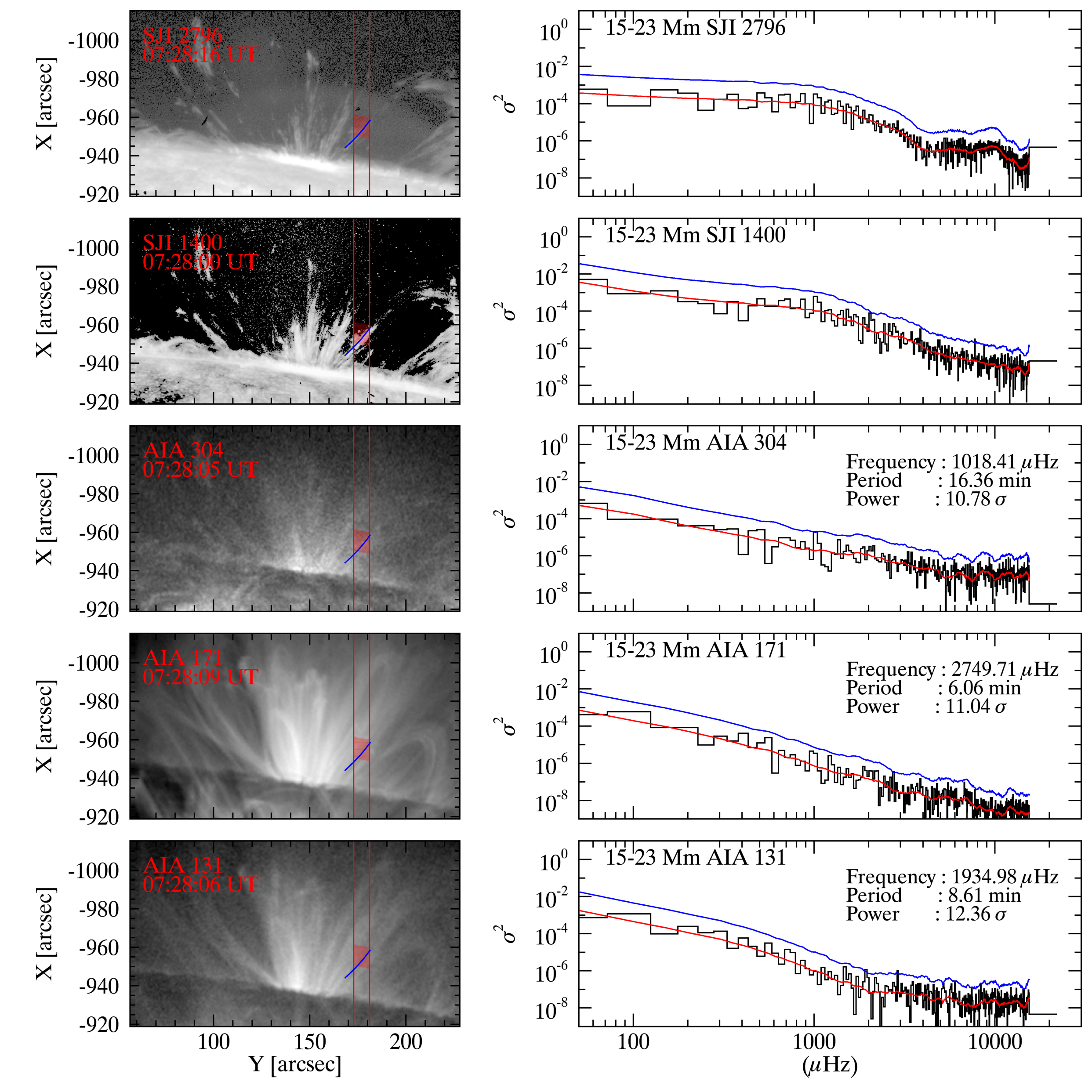}
\caption{Left columns: The SJI~2796~\AA, SJI~1400~\AA, AIA~304~\AA, AIA~171~\AA, and AIA~131~\AA\ images (from top to bottom). Right columns: Fourier power spectrum of the SJI~2796~\AA, SJI~1400~\AA, AIA~304~\AA, AIA~171~\AA, and AIA~131~\AA\ for the lightcurves constructed by summing over $15-23$~Mm heights (red squares in the left panels). The red and blue curves correspond to the average power and $10~\sigma$ power levels, respectively.\label{short_period}}
\end{figure*}

\begin{figure*}[ht]
\includegraphics[width=0.85\textwidth]{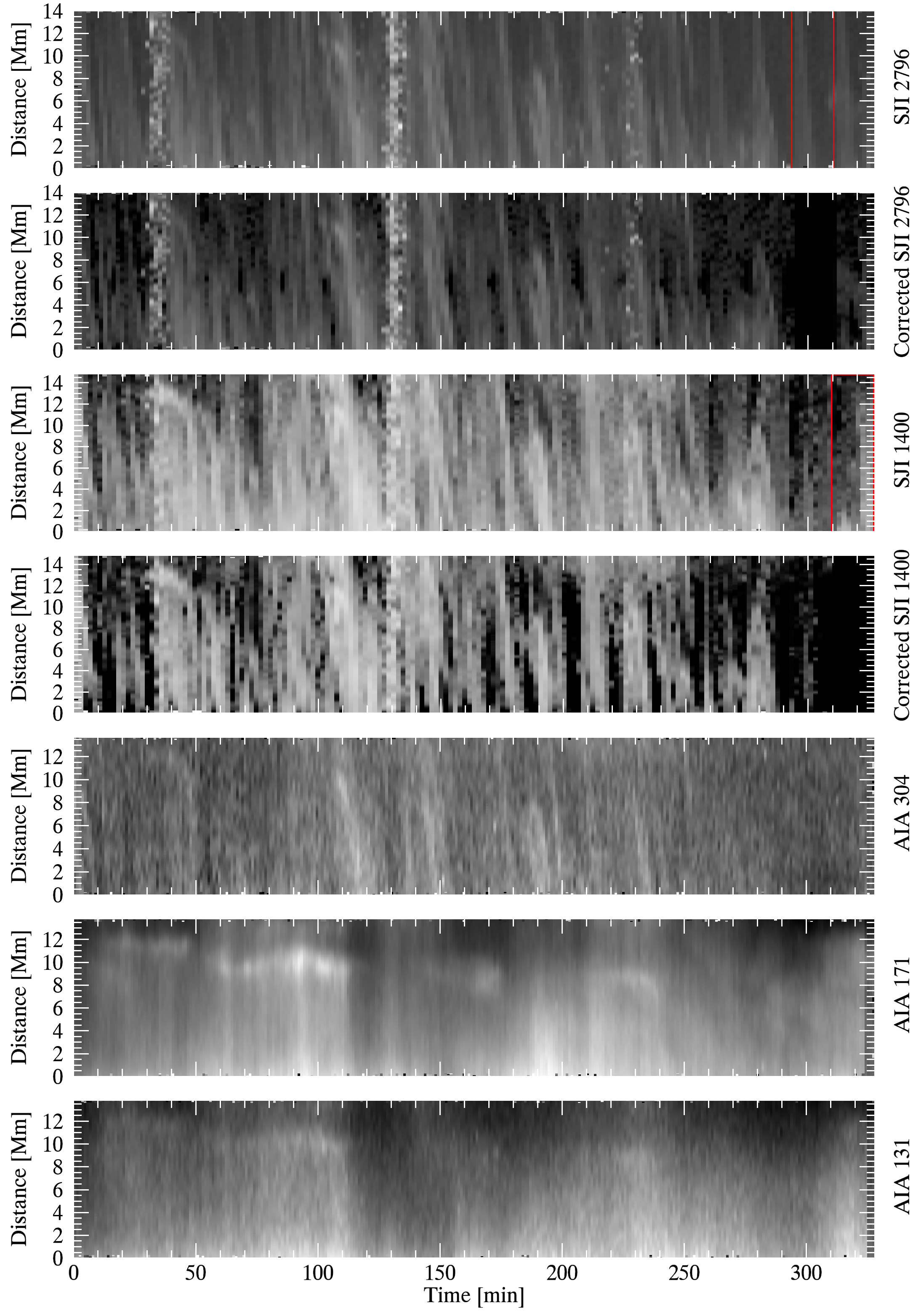}
\centering
\caption{Time-distance maps along the blue paths shown in Figure \ref{short_period} left panels for the five channels. The red rectangle area (on the first and third panel from the top) indicates the regions with the minimum solar activity that we selected to quantify the apparent motion variation effect for each IRIS SJI channel. The intensity in this time-distance area is subtracted from all the other rasters in this time-distance map to reduce the effects of the apparent motion. The original time-distance maps are given in the first and third panels from the top. The corrected time-distance maps are given in the 2nd and 4th panels. \label{short_period_td}}
\end{figure*}

In Figure~\ref{timdis}, there seem to be quasi-periodic downflows in the time-distance series over the entire field-of-view. To investigate this possibility and, more generally, the existence of short periodicity, we focus on the datasets and time range where rain is visible in both AIA and IRIS channels (i.e. 5.45~hr data in SJI~2796~\AA, SJI~1400~\AA, and AIA~304~\AA). We look for the existence of periodicities by first fixing a height above the limb and summing the intensities at that height both over the entire field-of-view and also locally by constraining the range in the Y axis. We construct lightcurves at every 8~Mm height starting from the solar limb and summing over a height thickness of 8~Mm. We then check for strong power peaks in the corresponding Fourier spectrum. In Figure~\ref{short_period}, we show a region at $15-23$~Mm heights (red square) in the SJI~2796~\AA, SJI~1400~\AA, AIA~304~\AA, AIA~171~\AA, and AIA~131~\AA\ images (left-columns) where a significant Fourier power is found (right columns). AIA~171~\AA\ has a peak of 11.04~$\sigma$, which corresponds to the frequency of 2749.71 $\mu$Hz (6.06~min). In AIA~131~\AA, two peaks of 12.36~$\sigma$ and 13.84~$\sigma$ powers were found at the frequency of 1934.98$\mu$Hz (8.61~min) and 2902.48$\mu$Hz (5.74~min). 
There is further no peak at exactly those frequencies in AIA~304~\AA. A peak of 10.78~$\sigma$ power was found at the frequency of detection of 1018.41$\mu$Hz (16.36~min) in the AIA~304~\AA\ channel. In the SJI~2796~\AA\ and SJI~1400~\AA, the power and confidence level at the same frequency are significantly lower (4.12$\sigma$, 6.26$\sigma$, respectively). As mentioned previously, due to the IRIS rastering mode, the apparent motion caused by the light scattering and imperfect flat-fielding introduces a non-uniform background noise. This effect periodically increases the intensity non-uniformly over the AR (with the centre part being more affected). The periodicity of this apparent motion is 17.2 min for both SJI~1400~\AA\ and SJI~2796~\AA\. Although the Fourier periodicity found is not exactly this one, it is possible that part of the power found in nearby frequencies is due to this apparent motion. To clear-up this possibility, in Figure \ref{short_period_td}, we show the time-distance map over the blue path shown in Figure \ref{short_period}, which was done with the help of the CRISPEX (CRisp SPectral EXplorer) tool \citep{Vissers_Rouppe_2012ApJ...750...22V}. To reduce the influence of the apparent motion effect, we first identify in the time-distance map the IRIS raster in each channel (1400 and 2796) with no (or minimal) coronal rain (indicated by the red rectangles in Figure \ref{short_period_td}). We then subtract the intensity in these rectangle areas from all the other IRIS rasters in the time-distance maps (see panels 2 and 4) in order to distinguish more clearly the coronal rain. The same short timescale rain showers can be clearly seen in SJI~1400~\AA\ and SJI~2796~\AA. These repetitive downflows are also seen in the time distance of the AIA~304~\AA, AIA~171~\AA, and AIA~131~\AA.

 
\section{Discussion and Conclusions \label{sec:sum}}

In an effort to further understand the fine structure of quiescent coronal rain,  we report in this study the first high-resolution statistical study of the quiescent coronal rain over an entire active region (AR) off-limb and over a significant time duration ($\approx 5.45$~hours). The duration of this observation is long enough to study the full cooling phase of the hours-long TNE cycles \citep{froment2015}. We aim to study the dynamics and morphology of coronal rain seen in various channels that probe different temperatures and densities, from its formation point high along the loops down to the chromosphere. Most, if not all, previous coronal rain observations are done either with ground-based telescopes, cases in which the rain off-limb can only be observed at the base of the loops (due to the inability of Adaptive Optics systems to operate off-limb), or focus on small rain events at much smaller scales compared to the AR scale. 
Coronal rain research is particularly relevant in the context of coronal heating since coronal rain properties can significantly constrain the spatio-temporal properties of the coronal heating mechanisms. 

Our statistical results show that rain clumps are seen at chromospheric and transition region temperatures corresponding to the formation ranges of \ion{Si}{IV} and \ion{Mg}{II} ions have widths from 0.2~Mm up to 2~Mm, and order of magnitude longer lengths, from a few Mm to 20~Mm. The results concerning the widths are very similar to those found by \citet{antolin2015} for a smaller rain event, despite the $2\times$ lower resolution of our dataset, and are also in agreement with simulation results \citep{Li2022ApJ...926..216L}. It is worth noting that these widths are roughly 5.8 times larger than those found in H$\alpha$ and 1.4 times larger in the same SJI channels of IRIS, which could be partly explained by the $6\times$ lower resolution of our dataset compared to the SST \citep{antolin2015} and the twice lower resolution compared to the IRIS dataset used in that study. Besides resolution, the different (larger) opacities in the \ion{Mg}{II} and \ion{Si}{IV} lines compared to H$\alpha$ observations from the ground likely contribute to the larger widths. Furthermore, the degradation of the SJI channels may also substantially contribute to increasing the rain widths. On the other hand, the lengths that we find are significantly longer, which is likely due to the much larger spatial scales analysed here and the fact that lengths are subject to longitudinal effects along loops (such as gas pressure and gravity). This is supported by the fact that the lengths found in AIA~304~\AA\ are similar to those found in \citet{antolin2015} with the same channel. Our width results suggest that factors defining coronal rain widths operate rapidly (as soon as it forms) and are rather constant along loop lengths (during their fall). The width of a coronal rain clump has been attributed on one hand to the fundamental scale over which coronal heating operates \citep{Antolin2022A}, which may itself be linked to the granular scales and associated currents \citep{Martinez2018}. Another possibility is that the rain width is set by the cooling mechanism (TI), through e.g. the wavelength of the unstable mode \citep{VanderLinden_1991SoPh..134..247V}. Even though the dataset provides good resolution for scales larger than 800 km, we acknowledge that the resolution may not be sufficient to capture structures at smaller scales, such as those around 500-600 km. However, we note that the Gaussian fitting technique enables us to confidently detect consistent variations in the widths around 800 km with height. We also acknowledge that low resolution and degradation effects may limit our ability to see a clear trend if present. In any case, this result would suggest that the mechanism (heating or cooling) does not vary in such a way as to modify the width of the clumps as they fall. 

On the dynamical side, rain clumps have a broad velocity distribution varying from a few km~s$^{-1}$ to $>100$ km~s$^{-1}$, which is in agreement with both observations \citep{de_groof2005,muller2005,antolin2012} and simulations \citep{fang2015,Li2022ApJ...926..216L}. 
We also checked the dynamical change of the rain with height. We found small downward accelerations at higher heights (between 30-40~Mm). However, during most of the fall, the rain shows almost constant velocities, with only a small linear increase on average. This linear increase or constant velocities at lower heights may be a combination of the effective gravity and pressure restructuring \citep{antolin2010,Oliver2014,Martinez-Gomez2020}. However, we did not find any higher velocities in the chromospheric SJI~2796~\AA\ emission, which would be expected due to the higher densities compared with the SJI~1400~\AA\ emission. This may be due to the strong multi-thermal character of the rain, as commented further below. 

Besides the downward motion, coronal rain is also known for exhibiting upward motions and complex changes in trajectories. However, the dynamical aspect of the upward motion of coronal rain has received limited attention in the current literature. It is worth highlighting those earlier studies, such as Antolin et al. (2010), has already noted changes in the trajectories of coronal rain, including the occurrence of upflows. In this study, we calculated the proportion of downflow to upflow motions of the clumps over the observational time, finding a ratio of 7 on average and a ubiquitous spatial occurrence. The ubiquitous upward motion of the coronal rain has also been obtained in a 2.5-D MHD simulation study by \citet{Li2022ApJ...926..216L}. We also found that the upflow speeds of coronal rain are on the same order as the downflows, with higher upward speeds than downward speeds on average. One possibility is that some upflows correspond to an apparent effect due to the change of opacity during the fall. Indeed, coronal rain, as it falls, undergoes cooling, resulting in an increase in opacity during the fall. This appearance can be sudden, leading to apparent fast upward motion. On the other hand, these dynamics may be real and reflect a significant role of gas pressure. Indeed, \citet{Mackay2001SoPh..198..289M} point out the fact that the chromosphere acts as a piston that can effectively stop the rain from falling due to the compression downstream. \citet{Adrover2021A&A...649A.142A} have analytically investigated the critical points along the loop (seen as a dynamical system) produced by these line tying conditions in the lower atmosphere, which leads to oscillations and therefore upflows as well. \citet{Oliver2014} have shown that gas pressure gradients effectively eliminate the downward acceleration. Therefore, gas pressure changes can also as easily lead to upflows with similar speeds as for the downflows. We observe the upward speeds to increase for decreasing height, which is in agreement with the larger gas pressure variation due to compression as the rain falls. Furthermore, the majority of the downflows we observe go towards the sunspot, with little downflows falling into adjacent regions, where the other loop footpoints are expected. This strongly suggests that the lower gas pressure expected within the sunspot (due to the strong magnetic field strength) produces siphon flow-type conditions, which become thermally unstable under the TNE-TI scenario. Many other observations of sunspots have also observed the rain prevalently going towards the umbra or penumbra \citep{Ahn2014SoPh..289.4117A, Kleint2014ApJ...789L..42K,antolin2015,Schad2016ApJ...833....5S,froment2020}. The impact from the rain is expected to produce strong rebound shocks and flows, which may constitute the strong pressure changes needed to modify the dynamics of trailing rain clumps and produce upflows.

We observe coronal rain distributed evenly to some extent in the POS above the AR (from one side and the other of the sunspot), which suggests that a very large number of loops exhibit coronal rain at some point during the 5.45~hr duration of the observation, irrespective of their geometry and footpoint location. An estimation of the volume involved has been done by \citet{Sahin2022ApJ...931L..27S} by quantifying the spatial and temporal extent of rain showers, which helps identify the coronal loops against the line-of-sight superposition due to optically thin coronal radiation. It was found that the total number of showers may be up to 150 with an estimated TNE volume of at least 50~\% the AR volume, suggesting the prevalence of TNE over the AR. This is in apparent contradiction with the results of \citet{Mikic_2013ApJ...773...94M,Froment2017ApJ...835..272F}, who show that loop geometry and heating asymmetries between footpoints play an important role in coronal rain formation and that small variation of these parameters can easily prevent TNE. However, \citet{Froment2018ApJ...855...52F,Pelouze_2022AA...658A..71P} find that such stringent conditions are relaxed for stronger footpoint heating, case in which TNE becomes prevalent in the parameter space. This would therefore suggest that the loops we observe are subject to relatively strong, stratified, and high-frequency heating. 

We also found that chromospheric emission in SJI~2796~\AA\ from coronal rain is highly correlated with the transition region emission in SJI~1400~\AA\ and AIA~304~\AA, indicating a strongly multi-thermal character, as previously suggested for isolated rain events \citep{antolin2015}, and supported by 2.5D radiative MHD simulations \citep{Antolin2022A}. This also indicates that the catastrophic cooling driven by thermal instability is very fast, occurring on timescales much faster than the free-fall time of the plasma, thereby leading to `complete' condensations rather than `incomplete' or `aborted' condensations for this active region \citep{Mikic_2013ApJ...773...94M}. 
However, the numerical results indicate much longer coronal rain lengths in transition region lines such as \ion{Si}{IV}~1402~\AA\ or \ion{He}{II}~304~\AA\ due to an extended tail at transition region temperatures, which we find in some isolated cases \citep{Antolin2022A}, but not an average. Accordingly, the chromospheric emission is only confined to the head of the rain in the simulation results. This suggests that thermal instability is actually far more extended along the loops, probably indicating excessive Joule heating in these numerical simulations.

We also found multiple long-period EUV pulsations associated with the observed coronal rain, all located at the top of the loops and with periods between 3 and 9 hours, strengthening the result of \citet{froment2020} that the two phenomena are 2 sides of the same coin, linked through the TNE-TI scenario. As stated in that paper, it is likely that we are only observing the tip-of-the-iceberg, with only the strongly periodic pulsations being detected with the Fourier method. Similarly, it is likely that we are only observing part of the rain associated with the pulsations that have periods longer than the duration of our dataset (for example, those with periods closer to 9 hr). As mentioned in \citet{Sahin2022ApJ...931L..27S}, this suggests that the rain and associated TNE-TI volume we observe above this active region is a lower limit. The fact that we observe them near the loop apexes may be due to the reduced superposition along the LOS, as also found in \citet{auchere2018} and \citet{froment2020}.

The TNE cycles are usually periods of hours, a timescale that is set by the radiative cooling time of a coronal loop. However,  in this study, we also observed much faster time scales (in the range of $5-20~$min), which makes it a challenge to explain in the TNE-TI scenario. This behaviour may partially correspond to a TNE cycle, in which the loop is continuously sustained by a siphon flow and only one leg of the loop is affected by TNE. In this case, the loop would never be fully depleted and temperatures are always at warm upper TR level. In such conditions, the radiative cooling timescale is always short and thermal stability may be triggered in a cyclic manner. Future modelling should explore the TNE parameter space to explore this possibility. 



\begin{acknowledgments}
P.A. acknowledges funding from his STFC Ernest Rutherford Fellowship (No. ST/R004285/2). IRIS is a NASA small explorer mission developed and operated by LMSAL, with mission operations executed at NASA Ames Research Center and major contributions to downlink communications funded by ESA and the Norwegian Space Centre. SDO is a mission for NASA’s Living With a Star (LWS) program. AIA is an instrument onboard the Solar Dynamics Observatory. All SDO data used in this work are available from the Joint Science Operations Center (http://jsoc.stanford.edu) without restriction. The National Solar Observatory (NSO) is operated by the Association of Universities for Research in Astronomy, Inc. (AURA), under cooperative agreement with the National Science Foundation. S.S., P.A, and C.F. were supported by the Programme National PNST of CNRS/INSU co-funded by CNES and CEA. This research was supported by the International Space Science Institute (ISSI) in Bern, through ISSI International Team project $\#545$ (``Observe Observe Local Think Global: What Solar Observations can Teach us about Multiphase Plasmas across Physical Scales”). We are grateful to the anonymous referee for the careful review and thoughtful comments, which greatly enhanced the clarity and rigour of this work.
\end{acknowledgments}

\bibliography{sample631}{}
\bibliographystyle{aasjournal}



\end{document}